\documentclass[reprint,superscriptaddress,amssymb,amsmath,aps,showpacs,10pt,floatfix,bibliography,pra]{revtex4-2}
\usepackage{graphicx}
\graphicspath{{figures/}}
\usepackage{color}
\usepackage[colorlinks,bookmarks=false,citecolor=darkblue,linkcolor=red,urlcolor=blue]{hyperref}

\definecolor{darkred}{rgb}{0.7,0.0,0.0}
\definecolor{darkblue}{rgb}{0,0.02,0.45}

\begin{document}

\begin{abstract}
The phonon renormalization across the semiconductor-to-metal crossover in FeSi is investigated by inelastic neutron scattering combined with \textit{ab-initio} lattice dynamical calculations. A significant part of reciprocal space with a particular focus on the 110$-$001 scattering plane is mapped by the time-of-flight inelastic neutron scattering data  taken below and above the crossover. Individual momentum values are investigated in more detail as a function of temperature. The data reveal that the anomalous phonon softening upon metallization is not exclusive to the high symmetry $R$ and $\Gamma$ points. Several other phonon modes around the $R$-point as well as the phonon modes at the $M$ and $X$ points of the Brillouin zone exhibit anomalous phonon softening with magnitudes comparable to that observed at the $R$-point. The momentum dependence of the phonon softening is reproduced by the lattice dynamical calculation based on the density functional perturbation theory. We discuss our findings with respect to the nature of the semiconductor-to-metal crossover in FeSi, for which different microscopic origins have been prooposed, i.e., lattice thermal disorder and electronic correlation effects.            

\end{abstract}
 ------------------------------------------------------------------------------------------------

\title{A combined inelastic neutron scattering and \textit{ab initio} lattice dynamics study of FeSi}

\author{Nazir Khan}
\email{nazirkhan91@gmail.com}
\affiliation{Institute for Quantum Materials and Technologies, Karlsruhe Institute of Technology, D-76021 Karlsruhe, Germany}

\author{Sven Krannich}
\affiliation{Institute for Solid State Physics, Karlsruhe Institute of Technology, D-76021 Karlsruhe, Germany}

\author{Dominic Boll}
\affiliation{Institute for Solid State Physics, Karlsruhe Institute of Technology, D-76021 Karlsruhe, Germany}

\author{Rolf Heid}
\affiliation{Institute for Quantum Materials and Technologies, Karlsruhe Institute of Technology, D-76021 Karlsruhe, Germany}

\author{Daniel Lamago}
\affiliation{Institute for Solid State Physics, Karlsruhe Institute of Technology, D-76021 Karlsruhe, Germany}
\affiliation{Laboratoire Léon Brillouin (CEA-CNRS), CEA Saclay, F-91911 Gif-sur-Yvette, France}

\author{A. Ivanov}
\affiliation{Institute Laue-Langevin, BP 156, 38042 Grenoble, France}

\author{David Voneshen}
\affiliation{ISIS Facility, Rutherford Appleton Laboratory, Chilton, Didcot, Oxfordshire OX11 0QX, United Kingdom}
\affiliation{Department of Physics, Royal Holloway University of London, TW20 0EX, UK}

\author{Frank Weber}
\email{frank.weber@kit.edu}
\affiliation{Institute for Quantum Materials and Technologies, Karlsruhe Institute of Technology, D-76021 Karlsruhe, Germany}

\date{\today}

\maketitle


\section{Introduction}
The family of transition-metal mono-silicides (TM-Si) Mn$_{1-x}$Fe$_{x-y}$Co$_y$Si crystallizing in the noncentrosymmetric cubic B20 structure hosts a very rich phase diagram showcasing a plethora of complex phenomena that are of interest for fundamental physics and applied science \cite{Pfleiderer2007,Schulz2012,Manyala2012,Pappas2021,Ban2018,Bann2018}. The phase diagram \cite{Manyala2004} as function of chemical substitution ($x$,$y$) shows a wide range of competing ground states with helimagnetic metallic magnetism (HMM), paramagnetic metallic (PMM) and paramagnetic insulating regions (PMI) as well as a magnetic quantum critical point \citep{Pappas2021}. Exotic phases such as partial order under hydrostatic pressure \cite{Pfleiderer2004} and skyrmion lattices in magnetic fields have been discovered in the end member MnSi \cite{Muhlbauer2009}. On the other hand, FeSi has the only insulating ground state and has been the focus of intense research interests for more than half a century due to its unusual temperature-dependent electrical, magnetic, and lattice dynamical properties~\cite{Jaccarino1967,Wertheim1965,Mandrus1995}.\\
\indent The $d$-electron compound FeSi is regarded as a correlated narrow-gap insulator and shows a remarkable similarity to $f$-electron Kondo insulators. The true nature of the electronic ground state is not yet unambigously settled. Experiments reveal a small charge gap $\Delta \sim$ 50-70 meV at low temperatures \cite{Jaccarino1967,Schlesinger1993,Degiorgi1994} and a cross-over to a (bad) metal at temperatures (100 K - 300 K) much smaller than $\Delta$ is observed in transport \cite{Wolfe1965,Buschinger1997,Paschen1997} and optical spectroscopy \cite{Schlesinger1993,vanderMarel1998,Degiorgi1994,Paschen1997,Damascelli1997,Menzel2009}, indicating that the charge gap is temperature dependent. Recently, size-dependent resistivity measurements on high quality single crystals of FeSi have provided evidence for a conducting surface state below 19 K \cite{Fang2018} which is reminiscent of a topological Kondo insulator. The insulator-to-metal crossover at room temperature is accompanied by a temperature-activated magnetism with a paramagnetic moment of $\sim$2 $\mu_B$ per Fe atom at $T\geq $500 K \cite{Jaccarino1967,Mandrus1995,Takagi1981}. Strongly temperature-dependent lattice dynamics in FeSi have been proposed to originate from metallization \cite{Delaire2011} and to be magnetically induced \cite{Krannich2015}.\\
\indent Theoretical models incorporating strong correlation effects such as local Coulomb interactions \cite{Tomczak2012}, lattice thermal disorder \cite{Delaire2011,Jarlborg1999}, and the orbital hybridization
between Fe 3$d$ and Si 2$p$ bands \cite{Mazurenko2010,Kune2008} have been invoked to explain the anomalous properties of FeSi. The observed phonon softening in FeSi has been captured by an earlier study based on finite-temperature first-principles calculations including thermal disorder effects\cite{Delaire2011}.  However, in a recent study \cite{Tomczak2012} based on realistic many-body calculations argues that electronic Coulomb correlations alone can quantitatively reproduce the signatures of the temperature induced cross-over in various observables: the spectral function, the optical conductivity, the spin susceptibility, and the Seebeck coefficient.\\
\indent Here, we address  the temperature-induced anomalous phonon renormalization in FeSi by means of  \textit{ab initio} lattice dynamical calculations and comprehensive single-crystal inelastic neutron scattering experiments. Going beyond previous studies, we investigate the momentum dependence of phonon renormalization at various high-symmetry points as well as in the vicinity of the $R$ point. We find that our quasi-harmonic  calculations for insulating FeSi and quasi-metallic FeSi resonably-well describe the observed phonon softening over a large momentum range.\\
\indent The manuscript is organized as follows. Section~\ref{sec:theory} contains theoretical details. In Sec.~\ref{sec:expt}, we present the details of different inelastic neutron scattering experiments performed on the FeSi single-crystal. Section~\ref{subsec:dfpt} describes the calculated phonon dispersion, electronic and phonon density of states, and temperature dependence of phonon energies for insulating and metallic FeSi. Section~\ref{subsec:inelas} reports the observed lattice dynamics based on inelastic neutron scattering experiments combined with theoretical calculations. In Section~\ref{sec:diss} we discuss our theoretical and experimental findings and compare them with earlier studies. The conclusions drawn in Section~\ref{sec:conclusion} brings to an end of the manuscript.

\section{Theory}
\label{sec:theory}
\textit{ab initio} lattice dynamical calculations based on density functional perturbation theory (DFPT) were performed in the framework of the mixed basis pseudopotential method \cite{Heid1999}. The exchange-correlation functional was treated in the local density approximation (LDA) in the Perdew–Wang parameterization \cite{Perdew1992}. Norm-conserving pseudopotentials for Fe and Si were constructed following the schema of Vanderbilt \cite{Vanderbilt1985} and include the Fe 3$s$ and 3$p$ semicore states in the valence space. The resulting deep potential can be efficiently treated in the mixed-basis scheme, which combines local functions together with plane waves for the representation of the valence states. Here, a combination of plane waves up to a kinetic energy of 22 Ry and local functions of $s$, $p$, and $d$ symmetry at the Fe sites gave sufficiently converged results. Brillouin zone integrations were performed with a cubic 8$\times$8$\times$8 $k-$point mesh (24 points in the irreducible Brillouin zone) in combination with a standard smearing technique using a Gaussian broadening of $\sigma$=0.05 eV and 0.2 eV. Dynamical matrices were calculated within DFPT on a cubic 4$\times$4$\times$4 $q$-point mesh, and then a standard Fourier interpolation technique was applied to get the full  phonon dispersion. Displayed results were obtained for the B20 structure ($P$2$_1$3) using experimental lattice constants and optimized internal parameters.
\section{EXPERIMENT}
\label{sec:expt}
\begin{figure*}[htb!]
	\centering
	\includegraphics[scale=0.7]{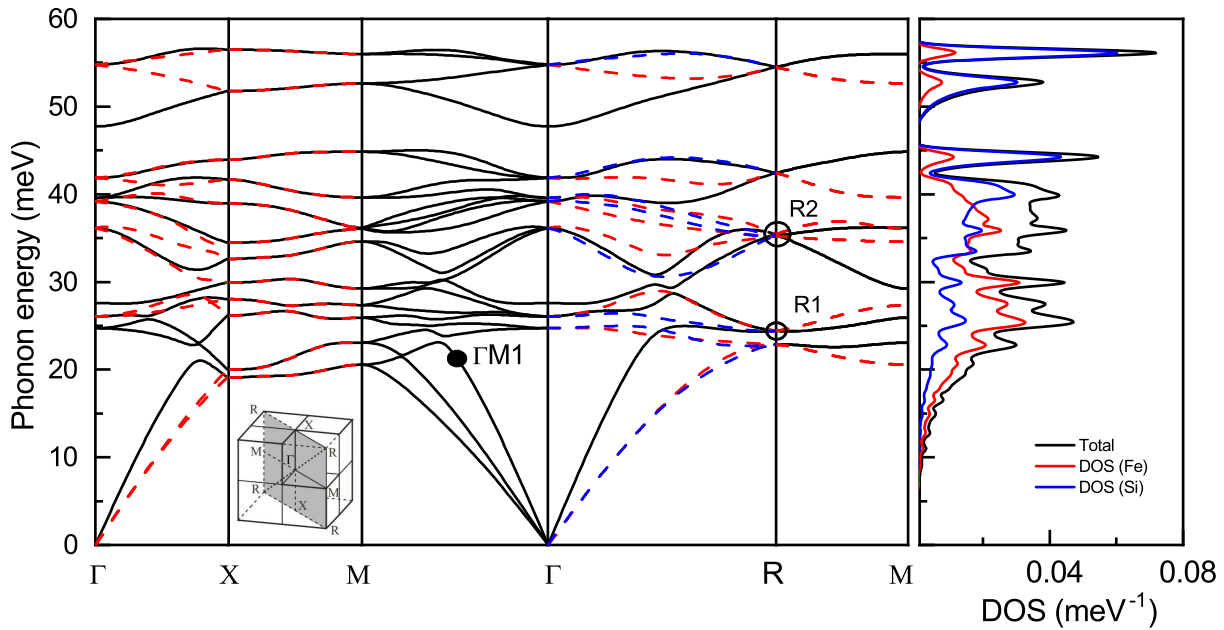}
	\caption{(left) Calculated phonon dispersion in FeSi. The phonons $R$1, $R$2
and   $\Gamma M$1 which we investigated also in a previous publication \cite{Krannich2015} are marked separately. Phonon dispersions are distinguished for their character. Solid and dashed lines refer to longitudnal and transverse characters, respectively. This distinciton was not possible along  $\Gamma M$. The inset shows the high-symmetry points of the Brillouin zone. (right) Calculated phonon density of states for FeSi (black curve) along with the partial densities of states for iron (red) and silicon (blue). \label{fig:bandphn}}
\end{figure*}
A high quality FeSi single crystal weighing $\sim$35 g was used for the inelastic neutron scattering (INS) experiments. INS measurements on the FeSi single crystal were performed on the direct-geometry, time-of-flight (TOF), chopper spectrometer MERLIN at the Rutherford Appleton Laboratory in Harwell, UK \cite{bewley2006}. The Merlin detectors cover scattering angles from $-45{}^{\circ}$ to +135${}^{\circ}$ horizontally and $\pm$30${}^{\circ}$ vertically. The sample was mounted on an Al sample holder using thin Al foil and Al sheet with (110)-(001) plane of the single crystal as the horizontal  scattering plane. The sample was then cooled by a closed-cycle refrigerator on the instrument. All data \cite{isisdata2020} were taken with an incident energy $E_i$ = 71.2 meV and covering the same volume of reciprocal space by rotating the sample over 70${}^{\circ}$ with angular steps of 0.25${}^{\circ}$. We used Gd chopper and a frequency of $f$ = 400 Hz was chosen to record data at $T$ = 10 K and 300 K. The raw data were corrected for detector efficiency by normalizing the intensities using a standard vanadium sample. Data was reduced using the Mantid package \cite{ARNOLD2014} and analyzed using HORACE \cite{EWINGS2016}. Selected phonon momenta were investigated in more detail on the thermal triple-axis spectrometer (TAS) 1T at the ORPHEE reactor at Laboratoire L\'{e}on Brillouin, CEA Saclay. Double-focusing copper (Cu 111) monochromator and  graphite (PG 002) analyser were used. The final energy at the analyzer was set to 14.7 meV allowing the use of a graphite filter to suppress higher-order scattering. The single crystal was mounted in a closed-cycle refrigerator allowing measurements in the temperature range 5 K $\leq$$T$$\leq$ 790 K. The magnetic field dependence of phonon renormalization for some selected phonon momenta was investigated  on the thermal triple-axis spectrometer IN8 at ILL, Grenoble using a 10 T vertical cryogenic magnet operating in the temperature range 2 K $\leq$$T$$\leq$ 300 K \cite{illdata2016}. Double-focusing copper (Cu 200) monochromator and analyser were used in the experiment, and a pyrolytic graphite (PG) was used as a filter with the final energy at the analyzer fixed to 14.7 meV. Scattering wave vectors \textbf{\textit{Q}} = {\boldmath$\tau$} + \textbf{\textit{q}} are expressed in reciprocal lattice units (r.l.u.) (2$\pi/a$,2$\pi/a$,2$\pi/a$) with the cubic lattice constant $a$. 

\section{Results}
\label{sec:results}
\subsection{Density functional perturbation theory}
\label{subsec:dfpt}
Figure~\ref{fig:bandphn} (left) shows calculated phonon dispersions in FeSi along different high-symmetry directions through the $\Gamma$, $X$, $M$, and $R$ high-symmetry points of the Brillouin zone as shown in the inset. Different colors are used to differentiate between different symmetries of the phonon displacement patterns. The experimental lattice constant at $T$ = 10 K ($a$ = 4.4745\AA) was used for the calculation \cite{Vocadlo2002,Sales1994}. The internal parameters were relaxed to obtain a force-free structure. The resulting values for the optimized structure are $u$(Fe) = 0.134 and $u$(Si)=0.840 and agree well with experimental values \cite{Wartchow1997}. The dispersion shown is in good agreement with a previous report \cite{Zhao2009}. Dot and circles denote three particular modes ($\Gamma M$1, $R$1, $R$2) which were investigated by some of us in a previous report \cite{Krannich2015}. Figure~\ref{fig:bandphn} (right) shows the corresponding phonon density of states together with the partial density of states for the vibrations of the iron and silicon atoms. The low-energy part of the phonon spectrum is dominated by the atomic  vibrations of Fe, whereas the vibrations of the significantly lighter Si atoms dominate the high-energy part of the phonon spectrum. Additional calculations using optimized lattice constants with LDA and GGA (general gradient approximation) produced phonon frequencies about 5-10\% harder than those presented in Fig.~\ref{fig:bandphn}. The origin is that both approximations, i.e., LDA and GGA, yield smaller lattice constants than experimentally reported.
\begin{figure}[h]
	\centering
	\includegraphics[scale=0.48]{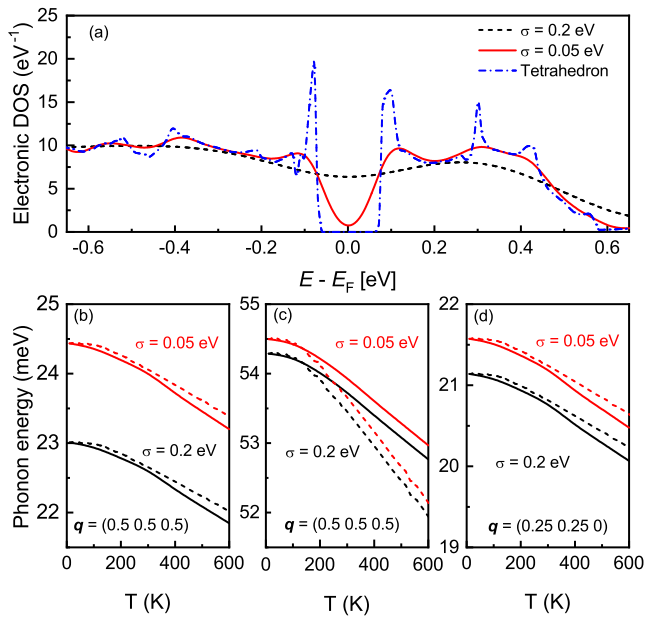}
	\caption{(a) Electronic density of states (eDOS) in the vicinity of the Fermi energy $E_F$ calculated within density-functional-theory (DFT) using a Gaussian smearing of $\sigma$ = 0.05 eV (solid line) and 0.2 eV (dashed line). eDOS calculated in the tetrahedron method (not compatible with the phonon calculations) is shown as well (dash-dotted line). (b) The calculated temperature dependence of the energy of the $R$1 phonon in the quasi-harmonic approximation, where temperature dependent lattice constants were taken from experimental reports \cite{Sales1994,Vocadlo2002}. The red (black) solid lines were calculated for $\sigma$ = 0.05 eV (0.2 eV).  (c, d) Analogous calculations for (c) a high energy mode at the $R$ point and (d) the phonon mode at $\textbf{\textit{q}}$ = (0.25,0.25,0), which is denoted as $\Gamma M$1 in Fig.~\ref{fig:bandphn}. The dashed lines in (b)-(d) correspond to the softening based on the reported thermal expansion \cite{Vocadlo2002} and a fixed Gr\"uneisen parameter of 2. \label{fig:eDOS}}
\end{figure}

~~FeSi has an insulating ground state and our calculations of the electronic structure show indeed a strong suppression of electronic states at $E_{\rm F}$, reminiscent of a gap in the excitation spectrum [Fig.~\ref{fig:eDOS}(a)], where a Gaussian smearing of $\sigma$ = 0.05 eV was used in the calculation. This value of $\sigma$ corresponds to the smallest value with which numerical convergence can be safely achieved. A more accurate description of the electronic density of states can be obtained by integrating over the electronic states using the tetrahedron method (see blue line in Fig.~\ref{fig:eDOS}(a)). In particular, the sharp maxima at the band edges, which were also observed in electron spectroscopy \cite{Arita2008,Klein2009}, are well reproduced in the tetrahedron method. However, lattice dynamical calculations in the used software package are not compatible with the tetrahedron method and, hence, are based on DFT calculations employing a Gaussian smearing. For the description of metallic FeSi at room temperature we used $\sigma$=0.2 eV. The resulting electronic density of states corresponds to the black curve in Fig.~\ref{fig:eDOS}(a) and shows strongly smeared structures. In particular, the density of states in the region of $E_F$ is significantly increased compared to the calculation with $\sigma$=0.05 eV. Thus, we can use different values of $\sigma$ to simulate effects on the lattice dynamics related to the closure of the electronic gap.

~~Although DFPT is a ground state calculation, changes in the lattice dynamics because of temperature dependent lattice constants can be studied within the quasi-harmonic  approximation (QH) \cite{Baroni2010,Debernardi2001,Quong1997}. For this purpose, the experimental lattice parameters \cite{Vocadlo2002} for temperatures 10 K $\leq$$T$$\leq$ 600 K were used in the calculation of phonon energies. The internal parameters were relaxed for each lattice constant. The calculation was performed using both $\sigma$=0.05 eV and $\sigma$=0.2 eV. Calculated temperature dependences of phonon energies are shown for three selected modes in Fig.~\ref{fig:eDOS}(b)-(d) (solid lines). These QH calculations are compared to an estimated softening based on the reported thermal expansion and  using a typical value of the Gr\"uneisen parameter of 2 [dashed lines in Figs.~\ref{fig:eDOS}(b)-(d))] \citep{Delaire2011}. In the QH calculations (solid lines) with a fixed value of $\sigma$, the softening of the phonons shown in Figs.~\ref{fig:eDOS} (b) and(d) is in good agreement with the estimates from the Gr\"uneisen model (dashed lines). However, for the high energy mode shown in Figs.~\ref{fig:eDOS}(c), the relative softening at high temperatures is overestimated in the Gr\"uneisen model compared to the QH calculations, which demonstrates that the QH calculations predict mode-dependent Gr\"uneisen parameters. The QH calculations reveal additional softening for $\sigma$=0.2 eV [black solid lines in Figs. 2(b)-(d)], i.e., for quasi- metallic FeSi. The relative softening due to the metallization is the highest ($\Delta E$=5.6\%) for the low energy phonon mode at $\textbf{\textit{q}}$=(0.5,0.5,0.5)[Fig.~\ref{fig:eDOS}(b)], while the high energy phonon mode at the same reduced $\textbf{\textit{q}}$ [Fig.~\ref{fig:eDOS}(c)] exhibits the lowest softening ($\Delta E$=0.4\%). The phonon mode at $\textbf{\textit{q}}$=(0.25,0.25,0) [Fig.~\ref{fig:eDOS}(d)] exhibits an intermediate value of the relative softening. Therefore, the calculations suggest that the phonon softening in FeSi due the metallization effect is strongly dependent on the energy and momentum of a phonon mode, which is further supported by our experimental results given below.  

\subsection{Experimental results}
\label{subsec:inelas}
\begin{figure}[h]
	\centering
	\includegraphics[scale=0.65]{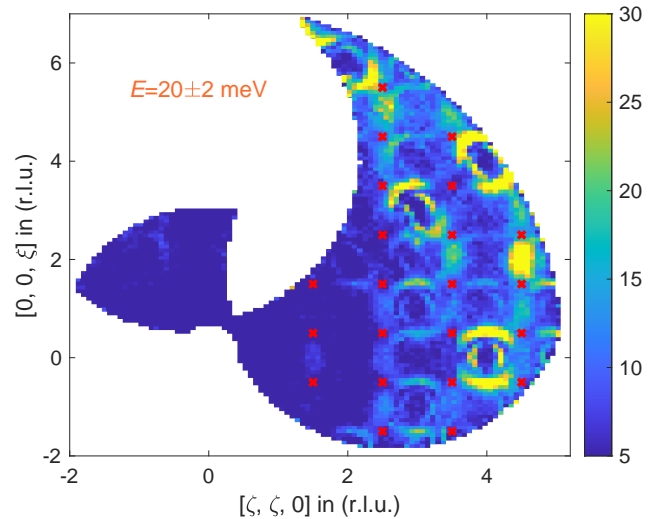}
	\caption{Horace plot of the inelastic neutron scattering data at  $T$=10 K with incident energy $E_i$=71.2 meV for the FeSi single crystal exhibiting the coverage in the [HHL] scattering plane for energy transfers binned over (20$\pm$2) meV. The red cross symbols represent the $R$ points in various Brillouin zones covered in our experiment.\label{fig:ecuts}}
\end{figure}
The coverage of momentum space in our TOF data for the horizontal [110]-[001] scattering plane is shown in Fig.~\ref{fig:ecuts} for energy transfers binned over 20$\pm$2 meV. The [111] direction is of particular interest, since anomalous phonon renormalization has been observed in FeSi for phonons at the $R$ point, $\textbf{\textit{q}}$ = (0.5,0.5,0.5), i.e., the zone boundary along the [111] direction \cite{Krannich2015}. The red-cross symbols in Fig.~\ref{fig:ecuts} indicate investigated $R$ points in Brillouin zones within the horizontal scattering plane, which are analyzed in more detail later in our report. 
\begin{figure}[h]
	\centering
	\includegraphics[scale=0.75]{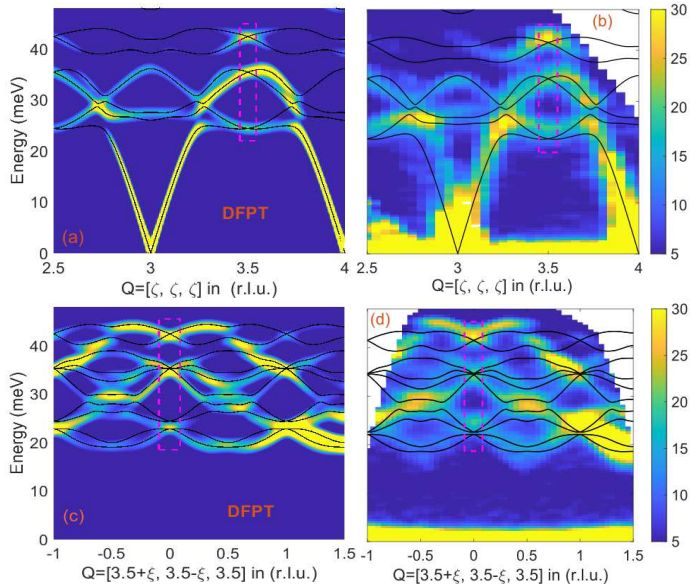}
	\caption{(a) Calculated phonon intensities overlaid with the corresponding dispersion lines along $\textbf{\textit{Q}}$ = 
[$\zeta$,$\zeta$,$\zeta$], 2.5 $\leq \zeta \leq$ 4, calculated with the reported lattice constants for $T$= 10 K and $\sigma$=0.05 eV. (b) Observed neutron scattering intensities for the same line in $\textbf{\textit{Q}}$ space at $T$=10 K. (c) and (d) show analogous results along the line $\textbf{\textit{Q}}$ = (3.5,3.5,3.5)+($\xi$,$-\xi$,0). Dashed rectangles in all panels indicate the position of the $R$ point at (3.5,3.5,3.5). \label{fig:dft-merlin}}
\end{figure}
\begin{figure}[h!tb]
	\centering
	\includegraphics[scale=0.5]{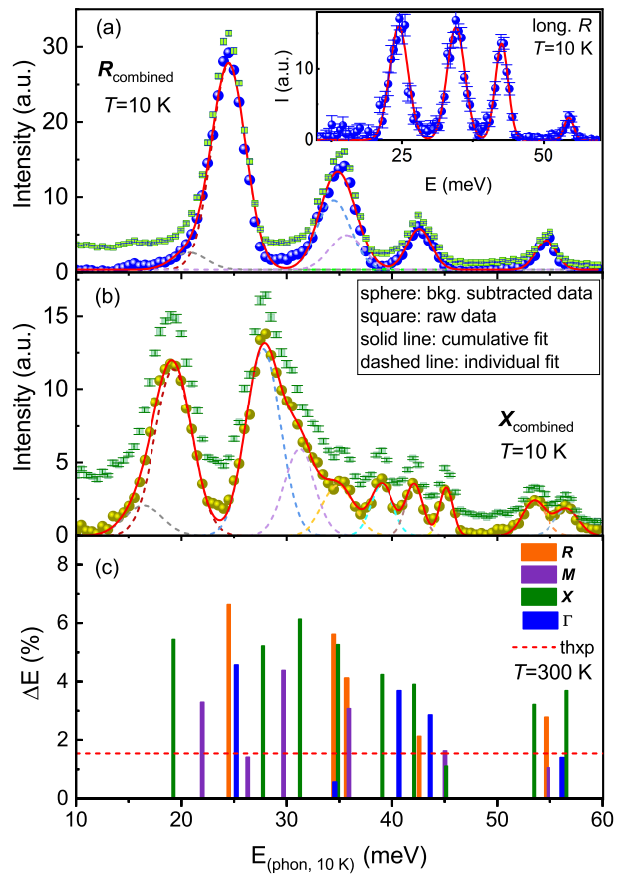}
	\caption{Analysis of combined (a) $R$ point and (b) $X$ point spectra. A linear background was subtracted from the raw data (squares). The resulting background subtracted data were approximated by a multi-peak fit (solid line) with the individual peaks shown as dashed lines. The number of peaks is in reasonable agreement with our calculations (see text). (c) Relative softening of phonon peaks on heating to $T$=300 K at the given high-symmetry points plotted as function of the low-temperature phonon energy. As a reference line, the dashed horizontal line indicates the softening because of the thermal expansion \cite{Vocadlo2002} assuming a fixed Gr\"uneisen parameter of $\gamma_{\textbf{\textit{k}},j}$ = 2 .\label{fig:esoften}}
\end{figure}

\indent We studied the dispersion of phonons going away from the $R$ point along the [111] and [1$\bar{1}$0] directions. Since previous data show strong renormalization of longitudinal phonons, we particularly investigated the region around $\textbf{\textit{Q}}$ = (3.5,3.5,3.5). There, we looked for anomalous phonon renormalization  on heating from 10 K to 300 K and compared our results to \textit{ab-initio} lattice dynamical calculations. We point out that the data obtained on MERLIN cover a large range of wave vectors. For instance, large phonon momenta near zone centers {\boldmath$\tau$} = (3,3,3) and (4,4,0)  [see Fig.~\ref{fig:ecuts}] can be accessed on triple-axis spectrometers in expense of the resolution or scattering intensity. Figure~\ref{fig:dft-merlin}(a) and (b) exhibit the calculated and experimentally observed phonon intensities at $T$=10 K, respectively, overlaid with the corresponding dispersion lines in the [$\zeta$,$\zeta$,$\zeta$] - Energy plane for 2.5 $\leq \zeta \leq$ 4 including $R$ points at half-integer values. The DFPT calculation shown in Fig.~\ref{fig:dft-merlin}(a) was performed using $\sigma$=0.05 eV and the experimental lattice parameter at 10 K \cite{Sales1994}. Figure~\ref{fig:dft-merlin}(c) and (d) are analogous plots, except that they exhibit dispersion along the line $\textbf{\textit{Q}}$ = (3.5,3.5,3.5)+($\xi$,$-\xi$,0). Overall calculation and experiment are in good agreement. We now look into the phonon renormalization on heating to room temperature in two ways: (1) combining data obtained at symmetry-equivalent wave vectors to improve statistics and (2) investigating phonons close to the $R$ point along the high symmetry directions shown in Fig.~\ref{fig:dft-merlin}.    
\subsubsection{Softening at high-symmetry points}
Despite the large size of the investigated FeSi single crystal, scattering intensities at individual absolute momenta $\textit{\textbf{Q}}$ are often not good enough to unambiguously approximate peak functions to reveal the temperature dependence of phonon energies. Therefore, we employed the following strategy for selected high-symmetry points: First, we obtained constant-$\textbf{\textit{Q}}$ scans for individual wave vectors from the $S(\textbf{\textit{Q}},\omega)$ datasets of FeSi collected at 10 K and 300 K using the Phonon Explorer software \cite{Dmitry2020} with a binning size ($\Delta H$=$\Delta K$=$\Delta L$) of $\pm$0.075 r.l.u. Then, data from symmetry-equivalent wave vectors was combined to improve the statistics.
The combined energy scans at 10 K for the $R$ and $X$ points are shown in Fig.~\ref{fig:esoften}(a) and Fig.~\ref{fig:esoften}(b), respectively. To determine the peak positions of the spectra, a linear sloping background was subtracted from the raw data sets [squares in Figs.~\ref{fig:esoften}(a) and (b)]. Gaussian functions  were approximated to the background subtracted data [spheres in Fig.~\ref{fig:esoften}(a) and Fig.~\ref{fig:esoften}(b)].
\begin{figure*}[htb]
	\centering
	\includegraphics[scale=0.65]{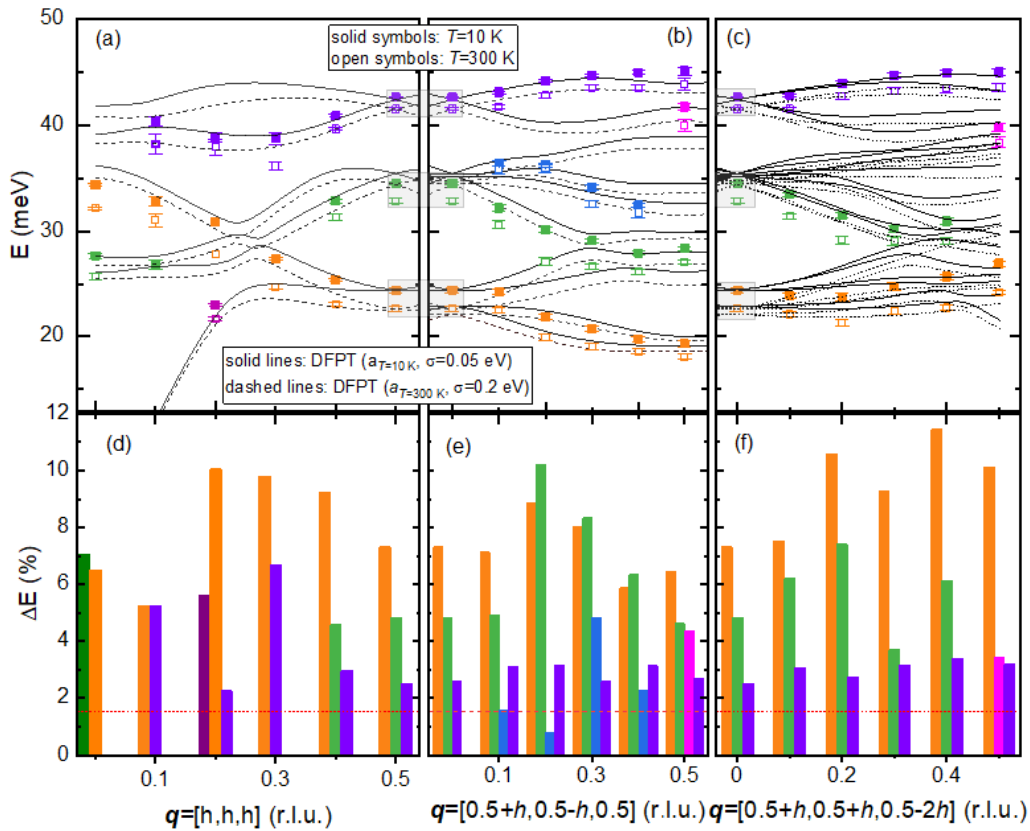}
	\caption{(a)-(c) Phonon dispersion in the vicinity of the $R$ point along three different directions, i.e., (a) $\delta \textbf{\textit{q}} \parallel$ [111], (b) $\delta \textbf{\textit{q}} \parallel$ [1$\bar{1}$0], and (c) $\delta \textbf{\textit{q}} \parallel$ [11$\bar{2}$]. Solid and dashed lines denote DFPT calculations for insulating  ($a_{T=10 K}$ = 4.55 \AA, $\sigma$ = 0.05 eV) and metallic ($a_{T=300 K}$ = 4.57\AA, $\sigma$ = 0.2 eV) FeSi, respectively, and the symbols are the corresponding experimental results. (d)-(f) Relative softening of the observed phonons as a function of phonon momentum transfer $\textbf{\textit{q}}$. As a reference line, the dashed lines indicate the softening due to the thermal expansion upon heating up to 300 K using a fixed Gr\"uneisen parameter of 2.\label{fig:qsoften}}
\end{figure*}
Starting with the $R$ point, we note that the high symmetry results in 4-fold degenerate phonon energies. Hence, there are only 6 different phonon energies. Our calculations show that for specific $R$ points with \textbf{\textit{q}} $\parallel$ {\boldmath$\tau$}, only four of these modes with well-separated energies have non-zero structure factors. We first identified these four modes from the spectra obtained by combining the energy scans at $\textbf{\textit{Q}}$=(1.5,1.5,1.5), $\textbf{\textit{Q}}$=(2.5,2.5,2.5), and $\textbf{\textit{Q}}$=(3.5,3.5,3.5) as shown in the inset of Fig.~\ref{fig:esoften}(a). As phonon energies remain the same for the same reduced $\textbf{\textit{q}}$=(0.5,0.5,0.5), the energies of the other two phonon modes in the combined energy scan [Fig.~\ref{fig:esoften}(a)] can now be determined following the prescription of DFPT. DFPT calculations suggest that the energies of these two modes lie close to the energies of the $R1$ and $R2$ modes identified from the spectrum of the phonons with longitudinal symmetry along the [111] direction [see Fig.~\ref{fig:bandphn} and inset of Fig.~\ref{fig:esoften}(a)]. Thus, a cumulative peak fitting to the combined energy scan [Fig.~\ref{fig:esoften}(a)] enabled the determination of all the six phonon energies for the $R$ point phonons. For the spectrum at other high symmetry points ($M$, $X$, and $\Gamma$) the situation is less favorable. Therefore, we considered only those peaks that are clearly visible in the spectra with significant intensities at both 10 K and 300 K for the cumulative peak fitting, e.g., the $X$ point shown in Fig.~\ref{fig:esoften}(b). The relative softening, $\Delta E$(\%), of different phonon modes at the $R$, $M$, $X$, and $\Gamma$ points of the Brillouin zone due to the crossover from the insulating state ($T$=10 K) to the metallic state ($T$=300 K) is shown in Fig.~\ref{fig:esoften}(c) together with the softening expected based on  thermal expansion and a fixed $\gamma_{\textbf{\textit{k}},j}$ = 2 as a reference line. Most of the phonon modes shown in Fig.~\ref{fig:esoften}(c) exhibit relative softening beyond this reference value ($\Delta E$=1.5\%) due to the thermal expansion upon heating up to 300 K. The columns in Fig.~\ref{fig:esoften}(c) demonstrate that several phonon modes at the $M$, $X$, and $\Gamma$ points exhibit strong softening  comparable to the anomalous softening of the $R$ point phonon modes reported previously [24]. We also see that the relative softening is stronger for lower energy phonons entailing larger involvement of the  heavier Fe atoms in lower-frequency vibrations (see  the partial phonon DOS plotted in Fig.~\ref{fig:bandphn}). Note that the relative softening obtained from the cumulative peak fitting for some phonons with very low intensities (and therefore low reliability) are not shown in Fig.~\ref{fig:esoften}(c).\\

\subsubsection{Softening around $R$ point}
~~After having focused on high-symmetry points in reciprocal space, we explore the phonon softening in FeSi in more detail as function of wave vector close to the $R$ point at $\textbf{\textit{Q}}$ = (3.5,3.5,3.5) [some data shown in Figs.~\ref{fig:dft-merlin}(b)(d)]. In Fig.~\ref{fig:qsoften}(a)-(c) we plot the observed dispersion going away from this $R$ point along [111], [1$\bar{1}$0] and [11$\bar{2}$] directions, respectively, at $T$=10 K (solid symbols) and $T$=300 K (open symbols). The experimental results are compared to the corresponding calculated dispersion line with $\sigma$ = 0.05 eV and 10 K lattice constants (solid lines) and $\sigma$ = 0.2 eV and room temperature lattice constants (dashed lines). There, we only show the dispersion lines of modes which have non-zero structure factor, and therefore, this number is different in panels (a-b-c) of Fig.~\ref{fig:qsoften}. The experimental phonon energies are determined from the  Gaussian fits to constant momentum lines extracted from the TOF data set, e.g., the data shown in Fig.~\ref{fig:dft-merlin}(b,d). The assignment of the observed peaks to the calculated phonon dispersion is comparatively straightforward for $\textbf{\textit{q}}$ = [$h$,$h$,$h$] since this is the highest symmetry direction in FeSi and only longitudinal phonon modes have non-zero intensities. The increasing number of phonon lines having non-vanishing structure factor along $\textbf{\textit{q}}$=[0.5+$h$,0.5-$h$,0.5] and [0.5+$h$,0.5+$h$,0.5-2$h$] allow only for a rough assignment of observed peaks to the calculated dispersion lines. Fig.~\ref{fig:qsoften}(d)-(f) exhibits the relative softening of the low temperature phonon frequencies in FeSi as a function of the phonon momentum transfer,  $\textbf{\textit{q}}$, upon metallization at $T$=300 K. All the observed low-lying phonons ($E<$ 35 meV) with fixed $\gamma_{k,j}$ = 2 [horizontal line in Figs.~\ref{fig:qsoften}(d)-(f)] exhibit softening beyond the estimate from thermal expansion and also comparable to that observed at the $R$ point ($\Delta E$=7\%) [grey shaded wave vectors in Figs.~\ref{fig:qsoften}(a)-(c)]. On average the softening of the high energy branches (purple) is significantly smaller than that for intermediate energies (green) and the softening is largest for the lowest energy phonons (orange). These observations manifest that the anomalous phonon renormalization is not exclusive to the phonons at the high symmetry points of the Brillouin zone.\\ 

\begin{figure*}[htb]
	\centering
	\includegraphics[scale=0.5]{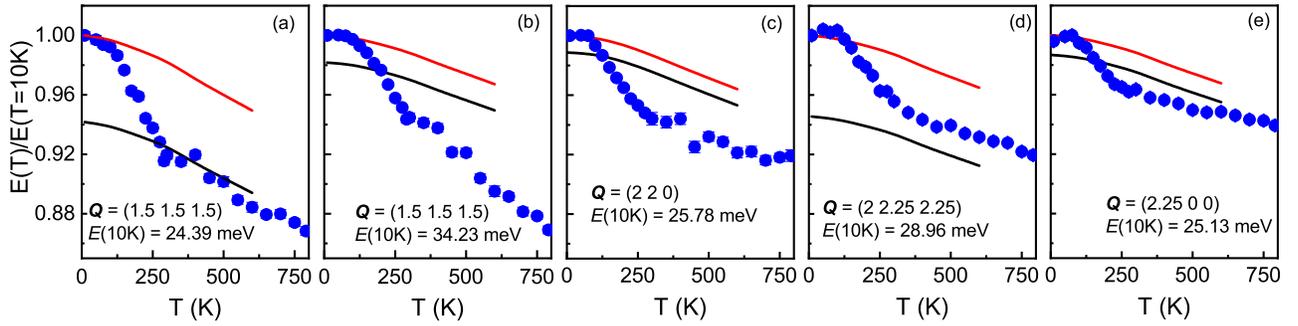}
	\caption{Temperature-dependent energies of selected phonon modes with energies up to 40 meV for temperatures 10 K $\leq T \leq$ 790 K. The phonon energies were normalized to the value at $T$ = 10 K. Red and black lines correspond to quasi-harmonic calculations with $\sigma$ = 0.05 eV and $\sigma$= 0.2 eV, respectively (see text).\label{fig:5qsoften}}
\end{figure*}
~~A detailed temperature dependence of softening for some selected phonon modes is investigated in the temperature range 10 K$ \leq T \leq $790 K ($E\leq$40 meV). In Figs.~\ref{fig:5qsoften}(a)-(e) we plot the variation of normalized phonon energies (symbols) obtained from the data collected at the thermal TAS 1T across the insulator-to-metal crossover temperature for five different phonon modes at a constant momentum transfer, $\textit{\textbf{Q}}$. The energies were normalized to the value at $T$ = 10 K. Red and black solid lines show the calculated phonon energies based on QH approximation for  $\sigma$= 0.05 eV and $\sigma$= 0.2 eV corresponding to an insulating- and metallic-type electronic DOS, respectively [see Fig.~\ref{fig:eDOS}(a)]. The data in Figs.~\ref{fig:5qsoften}(a)-(e) reveal that all the five phonon modes exhibit anomalous softening limited to the temperature range 100 K$ \leq T \leq $300 K. The slope of softening below 100 K and above 300 K agrees reasonably well with the QH calculations. Where our calculations show various degrees of agreement with the experimental observations, the results clearly demonstrate a difference in softening of up to a factor of 2 at room temperature, even for very similar phonon energies [see Figs.~\ref{fig:5qsoften}(a) and (e)].

\subsubsection{Phonon line brodening}
\begin{figure}[t!hb]
	\centering
	\includegraphics[scale=0.57]{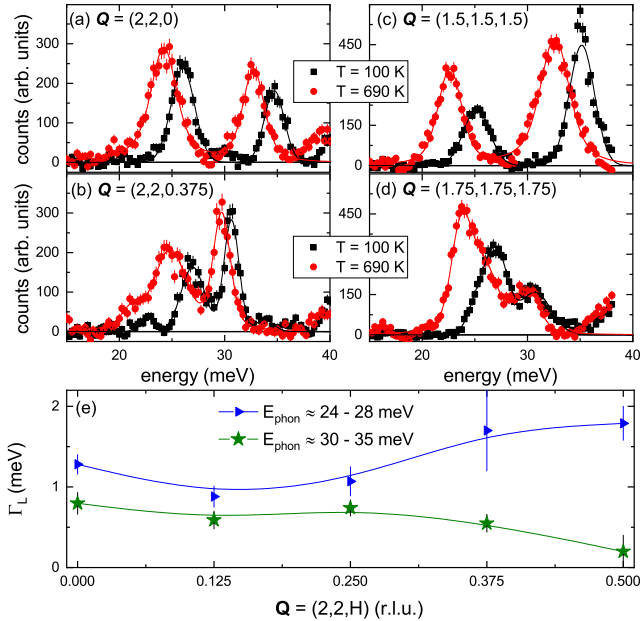}
	\caption{Phonon spectra  collected at $T$=100 K and 690 K in the vicinity of (a,b) the zone center {\boldmath$\tau$} = (2,2,0) and (c,d) the $R$ point at $\textbf{\textit{Q}}$=(1.5,1.5,1.5). For each spectrum a linear background was subtracted before peak fitting. Spectra at $T$= 690 K were fitted with Lorentz curves convoluted with the Gaussian resolution determined from the corresponding spectra at $T$= 100 K. (e) $\textbf{\textit{Q}}$-dependence of the approximated Lorentzian linewidth at the wave vectors $\textbf{\textit{Q}}$=(2,2,$h$).\label{fig:lw}}
\end{figure}
We also investigated the wave vector dependence of the broadening in zero field on the TAS 1T using a Cu (111) monochromator to improve the experimental resolution. Such an investigation is much more demanding than a mere determination of phonon energies. The approximated line width depends sensitively on the reliability of the estimated experimental background. Furthermore, phonons modes have to be well-separated in energy to allow an unambiguous analysis of the peak widths. 
Hence, we could only investigate the line width of phonon modes along the line $\textbf{\textit{Q}}$ = (2,2,$h$) with $h$ = 0-0.5 r.l.u. We approximated the data at $T$ = 100 K with Gaussians and used them as resolution function for further analysis. The high temperature data were analyzed by a Lorentzian convoluted with the Gaussian resolution determined at low temperature. Thus, we deduced the linewidth $\Gamma_L$ of the intrinsic phonon broadening at $T$ = 690 K which is shown in Fig.~\ref{fig:lw}(e). The observed phonon broadening along $\Gamma-X$ even increases for the mode around 25 meV but it practically vanishes for the mode above 30 meV.
~~The broadening of the phonon modes dispersing along the $\Gamma-R$, but away from $R$, cannot be determined unambiguously because the asymmetric lineshapes at $T$ = 690 K indicate strong overlap of multiple phonon peaks, e.g., at $\textbf{\textit{Q}}$=(1.75,1.75,1.75) [Fig.~\ref{fig:lw}(d)].\\

\subsubsection{Magnetic field dependence}
\begin{figure}[t!hb]
	\centering
	\includegraphics[scale=0.46]{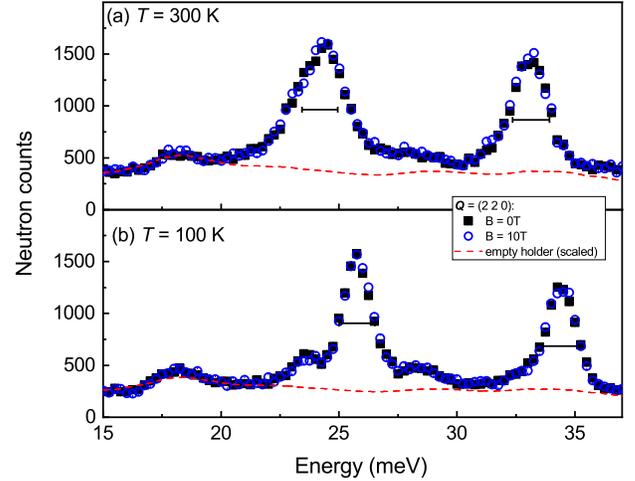}
	\caption{Phonon spectra at the zone center {\boldmath$\tau$} = (2,2,0) with and without applied magnetic field at (a) room temperature and (b) $T$=100 K. Horizontal bars indicate the width of the peak observed at low temperature. Dashed lines indicate the background taken from an empty can measurement for the same experimental configuration.   \label{fig:magns}}
\end{figure}
~~One hallmark of FeSi is the temperature-activated magnetism \cite{Jaccarino1967} where a large magnetic moment at high temperatures of up to 2 $\mu_B$ per Fe ion is suppressed at low temperatures $T \leq$ 100 K. Previously, we observed that phonons at the $R$ point are broadened with temperature following closely this temperature-activated behavior. Therefore, we performed INS measurements in an applied magnetic field of 10 T on the TAS IN8 to investigate a possible magnetic field dependence. We focused on the $R$ point at $\textbf{\textit{Q}}$ = (1.5,1.5,1.5) as well as the zone center at $\textbf{\textit{Q}}$ = (2,2,0) (Fig.~\ref{fig:magns}). While a structured background due to the complex sample environment prevents a detailed quantitative analysis, the raw data allow two conclusions: (1) A vertical magnetic field of 10 T has not detectable effect on the phonon peaks at either wave vector. (2) The previously reported broadening is clearly visible, e.g., at $\textbf{\textit{Q}}$ = (2,2,0) for the main peak around 25 meV (Fig.~\ref{fig:magns}).\\

\subsubsection{Results-overview}
In summary, our detailed INS investigation of the lattice dynamics of FeSi reveals:
(1) The anomalous softening in FeSi is limited in the temperature range 100 K$ \leq T \leq $300 K. The softening below 100 K and above 300 K can be approximated by the calculation of volume effect due to thermal expansion.
(2) The anomalous softening is widely distributed in $\textbf{\textit{Q}}$ though not all modes show it.
(3) Overall, the softening is stronger for phonons below 35 meV, which (according to PDOS) are Fe dominated.
(4) Broadening occurs also at various wave vectors - although only a few wave vectors could be checked due to the low symmetry and unclear selection rules in FeSi.
(5) Application of magnetic fields up to 10 T has no effect on the phonon properties of FeSi.
  
\section{Discussion}
\label{sec:diss}
The lattice dynamics in FeSi have been investigated before based on inelastic neutron scattering  \cite{Delaire2011,Krannich2015}, Raman and infrared spectroscopies \cite{Menzel2009,Racu2007}. The INS study addressed the phonon softening in FeSi based on the phonon DOS measured on a powder sample at a few selected temperatures and the single-crystal phonon dispersion taken at 10 K and 300 K. The Raman studies in FeSi mainly measured the energies of some zone-center phonons at 5 K and the temperature dependence of the phonon line width for a particular phonon mode ($E$ mode). Their experimental observations are in good agreement with our detailed study performed on a high-quality single crystal. From the experimental point of view, our study provides a deeper look into the anomalous phonon softening in FeSi. Unlike previous studies, our detailed temperature dependence of phonon energies (see Fig.~\ref{fig:5qsoften}) reveals clearly that the anomalous softening is only limited to the temperature window 100 K$ \leq T \leq $300 K. Additionally, based on the quantitative assessment we show that the relative softening  (Fig.~\ref{fig:esoften} and Figs.~\ref{fig:qsoften}) and broadening (Fig.~\ref{fig:lw}) are not exclusive to the high symmetry points$-$ rather it is widely distributed in $\textbf{\textit{Q}}$. Generally, we find that the magnitude of relative softening is strongest for phonons below 35 meV.  The calculation of partial phonon density of states [Fig.~\ref{fig:bandphn}] implies that the low energy phonons entail larger involvement of Fe atoms. As the conduction electrons result mainly from the Fe state, increased screening and, thus, reduced interatomic force-constant due to metallization could explain the stronger softening for the lower energy modes.\\
\indent Based on finite temperature molecular-dynamics (MD) calculations, anomalous temperature dependence of phonon in FeSi has been previously interpreted by considering the renormalization of the electronic structure by thermal disorder, leading to the closing of the narrow gap \cite{Delaire2011}. Though the simulated phonon density of states at different temperatures capture the phonon softening upon metallization, the MD  calculations could not reproduce correctly the temperature at which thermal disorder becomes important \cite{Delaire2011}.  Similarly, the closure of the electronic gap was predicted to happen at significantly higher temperatures \cite{Delaire2011}.\\
\indent A later study, based on a combination of density functional theory and dynamical mean-field theory (DMFT), proposed a new scenario in which FeSi is a band insulator at low temperatures and is metalized with increasing temperature through electronic correlation induced incoherence due to the unlocking of iron fluctuating moments \cite{Tomczak2012}. According to their theoretical picture, the cross-over to bad metallic behavior is not caused by a narrowing of the excitation gap, rather it is filled with incoherent weight that emerges with increasing temperature. This picture is validated by quantitative agreement with optical spectroscopy experiments which show spectral weight transfers over several electronvolts$-$ a common hallmark of correlation effects \cite{Rozenberg1996}.\\
\indent Thus, whether the thermal disorder effects are essential to simulate the lattice dynamics and spin fluctuations in FeSi at high temperatures, needs further investigation. In fact, our calculations for insulating FeSi and quasi-metallic FeSi, without considering any thermal disorder effects, appear to capture the observed phonon dispersion (Fig.~\ref{fig:dft-merlin} and Fig.~\ref{fig:qsoften}) and anomalous softening due to the metallization (Fig.~\ref{fig:5qsoften}) quite well. In molecular dynamics simulations \cite{Delaire2011}  which considers the thermal disorder effects, the gap $\Delta_{DFT}$ was shown to vanish abruptly for temperatures of the order of $\Delta_{DFT}/2$. This seems to be inconsistent with our experimental observation (Fig.~\ref{fig:5qsoften}) where we see that the anomalous phonon softening starts at around 100 K and ends at around 300 K, suggesting that the narrow gap in FeSi is gradually filled by electronic states and the filling is complete above 300 K. Such a gap closure in FeSi is consistent with the optical spectroscopy experiments \cite{Damascelli1997,Menzel2009} indicating the role of correlation effects in the electronic structure renormalization in FeSi. Thus, our study promotes that like other anomalous physical properties of FeSi, the microscopic origin behind  the anomalous lattice dynamics is also electronic in nature.\\ 
~~We also observed that broadening of the phonon linewidth is not limited only to the phonons at the high symmetry points [Fig.~\ref{fig:lw}(e)]. It appears that both phonon energy and linewidth exhibit strong renormalization over the entire Brillouin zone. The broadening at high temperatures could be attributed to a strong electron-phonon coupling as FeSi behaves like a metal as high temperatures. However, in our previous study \cite{Krannich2015} we observed that the phonon broadening  increases with increasing temperature beyond 300 K, and mimic the $T$-dependence of magnetization due to the development of a magnetic moment at the Fe site.  This implies that the strength of electron-phonon coupling increases with growing magnetism in FeSi at high temperatures, however, the presence of other interaction paths such as spin-phonon coupling can-not be ruled out.\\
\indent Additionally, we showed that a magnetic field of 10 T has no effect on the lattice dynamics even in the metallic phase where Fe develops a temperature induced local magnetic moment [Fig.~\ref{fig:magns}]. This is likely due to the mismatch in energy scales between the 10 T field and the spin fluctuations which are significant at very high temperatures.  Thus, it could be very interesting to see how the lattice dynamics in the isostructural metallic helimagnet MnSi respond to an external magnetic field. There, a magnetic field as low as 0.5 T can fully spin-polarize the system through an intermediate conical magnetic phase  \cite{Muhlbauer2009}.\\     

\section{Conclusion}
\label{sec:conclusion}
We have reported a comprehensive inelastic neutron scattering and \textit{ab} \textit{initio} theoretical investigation of the lattice dynamics of FeSi, focusing on the phonon renormalization across the temperature induced metal-to-insulator cross-over. The anomalous softening is found to be limited to the temperature range where FeSi gradually crosses over from a small-gap semiconductor to a bad metal. The softening is widely distributed in $\textbf{\textit{Q}}$ and appears stronger for phonons having energies below 35 meV i.e., vibrations dominated by Fe motions. Phonon broadening is also distributed in $\textbf{\textit{Q}}$ space. Overall, our theoretical calculations are in reasonable agreement with the experimental observations, which demonstrates that the anomalous phonon renormalization can also be reproduced without considering lattice thermal disorder effects. Apparently, magnetic field has no effects on the phonons in FeSi, however, we anticipate that lattice dynamics in the isostructural but magnetically ordered MnSi  may reveal some interesting field-dependent behavior. 
\acknowledgments
This work was funded by the Deutsche Forschungsgemeinschaft (DFG, German Research Foundation) under Project No. 419331252. Experiments at the ISIS Pulsed Neutron and Muon Source were supported by a beamtime allocation from the Science and Technology Facilities Council.
\bibliography{FeSi}

\begin{thebibliography}{48}%
\makeatletter
\providecommand \@ifxundefined [1]{%
 \@ifx{#1\undefined}
}%
\providecommand \@ifnum [1]{%
 \ifnum #1\expandafter \@firstoftwo
 \else \expandafter \@secondoftwo
 \fi
}%
\providecommand \@ifx [1]{%
 \ifx #1\expandafter \@firstoftwo
 \else \expandafter \@secondoftwo
 \fi
}%
\providecommand \natexlab [1]{#1}%
\providecommand \enquote  [1]{``#1''}%
\providecommand \bibnamefont  [1]{#1}%
\providecommand \bibfnamefont [1]{#1}%
\providecommand \citenamefont [1]{#1}%
\providecommand \href@noop [0]{\@secondoftwo}%
\providecommand \href [0]{\begingroup \@sanitize@url \@href}%
\providecommand \@href[1]{\@@startlink{#1}\@@href}%
\providecommand \@@href[1]{\endgroup#1\@@endlink}%
\providecommand \@sanitize@url [0]{\catcode `\\12\catcode `\$12\catcode
  `\&12\catcode `\#12\catcode `\^12\catcode `\_12\catcode `\%12\relax}%
\providecommand \@@startlink[1]{}%
\providecommand \@@endlink[0]{}%
\providecommand \url  [0]{\begingroup\@sanitize@url \@url }%
\providecommand \@url [1]{\endgroup\@href {#1}{\urlprefix }}%
\providecommand \urlprefix  [0]{URL }%
\providecommand \Eprint [0]{\href }%
\providecommand \doibase [0]{https://doi.org/}%
\providecommand \selectlanguage [0]{\@gobble}%
\providecommand \bibinfo  [0]{\@secondoftwo}%
\providecommand \bibfield  [0]{\@secondoftwo}%
\providecommand \translation [1]{[#1]}%
\providecommand \BibitemOpen [0]{}%
\providecommand \bibitemStop [0]{}%
\providecommand \bibitemNoStop [0]{.\EOS\space}%
\providecommand \EOS [0]{\spacefactor3000\relax}%
\providecommand \BibitemShut  [1]{\csname bibitem#1\endcsname}%
\let\auto@bib@innerbib\@empty
\bibitem [{\citenamefont {Pfleiderer}\ \emph {et~al.}(2007)\citenamefont
  {Pfleiderer}, \citenamefont {B\"oni}, \citenamefont {Keller}, \citenamefont
  {R\"ossler},\ and\ \citenamefont {Rosch}}]{Pfleiderer2007}%
  \BibitemOpen
  \bibfield  {author} {\bibinfo {author} {\bibfnamefont {C.}~\bibnamefont
  {Pfleiderer}}, \bibinfo {author} {\bibfnamefont {P.}~\bibnamefont {B\"oni}},
  \bibinfo {author} {\bibfnamefont {T.}~\bibnamefont {Keller}}, \bibinfo
  {author} {\bibfnamefont {U.~K.}\ \bibnamefont {R\"ossler}},\ and\ \bibinfo
  {author} {\bibfnamefont {A.}~\bibnamefont {Rosch}},\ }\bibfield  {title}
  {\bibinfo {title} {Non-fermi liquid metal without quantum criticality},\
  }\href {https://doi.org/10.1126/science.1142644} {\bibfield  {journal}
  {\bibinfo  {journal} {Science}\ }\textbf {\bibinfo {volume} {316}},\ \bibinfo
  {pages} {1871} (\bibinfo {year} {2007})}\BibitemShut {NoStop}%
\bibitem [{\citenamefont {Schulz}\ \emph {et~al.}(2012)\citenamefont {Schulz},
  \citenamefont {Ritz}, \citenamefont {Bauer}, \citenamefont {Halder},
  \citenamefont {Wagner}, \citenamefont {Franz}, \citenamefont {Pfleiderer},
  \citenamefont {Everschor}, \citenamefont {Garst},\ and\ \citenamefont
  {Rosch}}]{Schulz2012}%
  \BibitemOpen
  \bibfield  {author} {\bibinfo {author} {\bibfnamefont {T.}~\bibnamefont
  {Schulz}}, \bibinfo {author} {\bibfnamefont {R.}~\bibnamefont {Ritz}},
  \bibinfo {author} {\bibfnamefont {A.}~\bibnamefont {Bauer}}, \bibinfo
  {author} {\bibfnamefont {M.}~\bibnamefont {Halder}}, \bibinfo {author}
  {\bibfnamefont {M.}~\bibnamefont {Wagner}}, \bibinfo {author} {\bibfnamefont
  {C.}~\bibnamefont {Franz}}, \bibinfo {author} {\bibfnamefont
  {C.}~\bibnamefont {Pfleiderer}}, \bibinfo {author} {\bibfnamefont
  {K.}~\bibnamefont {Everschor}}, \bibinfo {author} {\bibfnamefont
  {M.}~\bibnamefont {Garst}},\ and\ \bibinfo {author} {\bibfnamefont
  {A.}~\bibnamefont {Rosch}},\ }\bibfield  {title} {\bibinfo {title} {Emergent
  electrodynamics of skyrmions in a chiral magnet},\ }\href
  {https://doi.org/10.1038/nphys2231} {\bibfield  {journal} {\bibinfo
  {journal} {Nature Physics}\ }\textbf {\bibinfo {volume} {8}},\ \bibinfo
  {pages} {301} (\bibinfo {year} {2012})}\BibitemShut {NoStop}%
\bibitem [{\citenamefont {Manyala}\ \emph {et~al.}(2012)\citenamefont
  {Manyala}, \citenamefont {Sidis}, \citenamefont {DiTusa}, \citenamefont
  {Aeppli}, \citenamefont {Young},\ and\ \citenamefont {Fisk}}]{Manyala2012}%
  \BibitemOpen
  \bibfield  {author} {\bibinfo {author} {\bibfnamefont {N.}~\bibnamefont
  {Manyala}}, \bibinfo {author} {\bibfnamefont {Y.}~\bibnamefont {Sidis}},
  \bibinfo {author} {\bibfnamefont {J.~F.}\ \bibnamefont {DiTusa}}, \bibinfo
  {author} {\bibfnamefont {G.}~\bibnamefont {Aeppli}}, \bibinfo {author}
  {\bibfnamefont {D.}~\bibnamefont {Young}},\ and\ \bibinfo {author}
  {\bibfnamefont {Z.}~\bibnamefont {Fisk}},\ }\bibfield  {title} {\bibinfo
  {title} {Magnetoresistance from quantum interference effects in
  ferromagnets},\ }\href {https://doi.org/10.1038/35007030} {\bibfield
  {journal} {\bibinfo  {journal} {Nature}\ }\textbf {\bibinfo {volume} {404}},\
  \bibinfo {pages} {581} (\bibinfo {year} {2012})}\BibitemShut {NoStop}%
\bibitem [{\citenamefont {Pappas}\ \emph {et~al.}(2021)\citenamefont {Pappas},
  \citenamefont {Leonov}, \citenamefont {Bannenberg}, \citenamefont {Fouquet},
  \citenamefont {Wolf},\ and\ \citenamefont {Weber}}]{Pappas2021}%
  \BibitemOpen
  \bibfield  {author} {\bibinfo {author} {\bibfnamefont {C.}~\bibnamefont
  {Pappas}}, \bibinfo {author} {\bibfnamefont {A.~O.}\ \bibnamefont {Leonov}},
  \bibinfo {author} {\bibfnamefont {L.~J.}\ \bibnamefont {Bannenberg}},
  \bibinfo {author} {\bibfnamefont {P.}~\bibnamefont {Fouquet}}, \bibinfo
  {author} {\bibfnamefont {T.}~\bibnamefont {Wolf}},\ and\ \bibinfo {author}
  {\bibfnamefont {F.}~\bibnamefont {Weber}},\ }\bibfield  {title} {\bibinfo
  {title} {Evolution of helimagnetic correlations when approaching the quantum
  critical point of {Mn$_{1-x}$Fe$_{x}$Si}},\ }\href
  {https://doi.org/10.1103/PhysRevResearch.3.013019} {\bibfield  {journal}
  {\bibinfo  {journal} {Phys. Rev. Research}\ }\textbf {\bibinfo {volume}
  {3}},\ \bibinfo {pages} {013019} (\bibinfo {year} {2021})}\BibitemShut
  {NoStop}%
\bibitem [{\citenamefont {Bannenberg}\ \emph
  {et~al.}(2018{\natexlab{a}})\citenamefont {Bannenberg}, \citenamefont
  {Weber}, \citenamefont {Lefering}, \citenamefont {Wolf},\ and\ \citenamefont
  {Pappas}}]{Ban2018}%
  \BibitemOpen
  \bibfield  {author} {\bibinfo {author} {\bibfnamefont {L.~J.}\ \bibnamefont
  {Bannenberg}}, \bibinfo {author} {\bibfnamefont {F.}~\bibnamefont {Weber}},
  \bibinfo {author} {\bibfnamefont {A.~J.~E.}\ \bibnamefont {Lefering}},
  \bibinfo {author} {\bibfnamefont {T.}~\bibnamefont {Wolf}},\ and\ \bibinfo
  {author} {\bibfnamefont {C.}~\bibnamefont {Pappas}},\ }\bibfield  {title}
  {\bibinfo {title} {Magnetization and ac susceptibility study of the cubic
  chiral magnet {Mn$_{1-x}$Fe$_x$Si}},\ }\href
  {https://doi.org/10.1103/PhysRevB.98.184430} {\bibfield  {journal} {\bibinfo
  {journal} {Phys. Rev. B}\ }\textbf {\bibinfo {volume} {98}},\ \bibinfo
  {pages} {184430} (\bibinfo {year} {2018}{\natexlab{a}})}\BibitemShut
  {NoStop}%
\bibitem [{\citenamefont {Bannenberg}\ \emph
  {et~al.}(2018{\natexlab{b}})\citenamefont {Bannenberg}, \citenamefont
  {Dalgliesh}, \citenamefont {Wolf}, \citenamefont {Weber},\ and\ \citenamefont
  {Pappas}}]{Bann2018}%
  \BibitemOpen
  \bibfield  {author} {\bibinfo {author} {\bibfnamefont {L.~J.}\ \bibnamefont
  {Bannenberg}}, \bibinfo {author} {\bibfnamefont {R.~M.}\ \bibnamefont
  {Dalgliesh}}, \bibinfo {author} {\bibfnamefont {T.}~\bibnamefont {Wolf}},
  \bibinfo {author} {\bibfnamefont {F.}~\bibnamefont {Weber}},\ and\ \bibinfo
  {author} {\bibfnamefont {C.}~\bibnamefont {Pappas}},\ }\bibfield  {title}
  {\bibinfo {title} {Evolution of helimagnetic correlations in
  {Mn$_{1-x}$Fe$_x$Si} with doping: A small-angle neutron scattering study},\
  }\href {https://doi.org/10.1103/PhysRevB.98.184431} {\bibfield  {journal}
  {\bibinfo  {journal} {Phys. Rev. B}\ }\textbf {\bibinfo {volume} {98}},\
  \bibinfo {pages} {184431} (\bibinfo {year} {2018}{\natexlab{b}})}\BibitemShut
  {NoStop}%
\bibitem [{\citenamefont {Manyala}\ \emph {et~al.}(2004)\citenamefont
  {Manyala}, \citenamefont {Sidis}, \citenamefont {DiTusa}, \citenamefont
  {Aeppli}, \citenamefont {Young},\ and\ \citenamefont {Fisk}}]{Manyala2004}%
  \BibitemOpen
  \bibfield  {author} {\bibinfo {author} {\bibfnamefont {N.}~\bibnamefont
  {Manyala}}, \bibinfo {author} {\bibfnamefont {Y.}~\bibnamefont {Sidis}},
  \bibinfo {author} {\bibfnamefont {J.~F.}\ \bibnamefont {DiTusa}}, \bibinfo
  {author} {\bibfnamefont {G.}~\bibnamefont {Aeppli}}, \bibinfo {author}
  {\bibfnamefont {D.~P.}\ \bibnamefont {Young}},\ and\ \bibinfo {author}
  {\bibfnamefont {Z.}~\bibnamefont {Fisk}},\ }\bibfield  {title} {\bibinfo
  {title} {Large anomalous hall effect in a silicon-based magnetic
  semiconductor},\ }\href {https://doi.org/10.1038/nmat1103} {\bibfield
  {journal} {\bibinfo  {journal} {Nature Materials}\ }\textbf {\bibinfo
  {volume} {3}},\ \bibinfo {pages} {255} (\bibinfo {year} {2004})}\BibitemShut
  {NoStop}%
\bibitem [{\citenamefont {Pfleiderer}\ \emph {et~al.}(2004)\citenamefont
  {Pfleiderer}, \citenamefont {Reznik}, \citenamefont {Pintschovius},
  \citenamefont {v.~L\"ohneysen}, \citenamefont {Garst},\ and\ \citenamefont
  {Rosch}}]{Pfleiderer2004}%
  \BibitemOpen
  \bibfield  {author} {\bibinfo {author} {\bibfnamefont {C.}~\bibnamefont
  {Pfleiderer}}, \bibinfo {author} {\bibfnamefont {D.}~\bibnamefont {Reznik}},
  \bibinfo {author} {\bibfnamefont {L.}~\bibnamefont {Pintschovius}}, \bibinfo
  {author} {\bibfnamefont {H.}~\bibnamefont {v.~L\"ohneysen}}, \bibinfo
  {author} {\bibfnamefont {M.}~\bibnamefont {Garst}},\ and\ \bibinfo {author}
  {\bibfnamefont {A.}~\bibnamefont {Rosch}},\ }\bibfield  {title} {\bibinfo
  {title} {Partial order in the non-fermi-liquid phase of {MnSi}},\ }\href
  {https://doi.org/10.1038/nature02232} {\bibfield  {journal} {\bibinfo
  {journal} {Nature}\ }\textbf {\bibinfo {volume} {427}},\ \bibinfo {pages}
  {227} (\bibinfo {year} {2004})}\BibitemShut {NoStop}%
\bibitem [{\citenamefont {M\"uhlbauer}\ \emph {et~al.}(2009)\citenamefont
  {M\"uhlbauer}, \citenamefont {Binz}, \citenamefont {Jonietz}, \citenamefont
  {Pfleiderer}, \citenamefont {Rosch}, \citenamefont {Neubauer}, \citenamefont
  {Georgii},\ and\ \citenamefont {B\"oni}}]{Muhlbauer2009}%
  \BibitemOpen
  \bibfield  {author} {\bibinfo {author} {\bibfnamefont {S.}~\bibnamefont
  {M\"uhlbauer}}, \bibinfo {author} {\bibfnamefont {B.}~\bibnamefont {Binz}},
  \bibinfo {author} {\bibfnamefont {F.}~\bibnamefont {Jonietz}}, \bibinfo
  {author} {\bibfnamefont {C.}~\bibnamefont {Pfleiderer}}, \bibinfo {author}
  {\bibfnamefont {A.}~\bibnamefont {Rosch}}, \bibinfo {author} {\bibfnamefont
  {A.}~\bibnamefont {Neubauer}}, \bibinfo {author} {\bibfnamefont
  {R.}~\bibnamefont {Georgii}},\ and\ \bibinfo {author} {\bibfnamefont
  {P.}~\bibnamefont {B\"oni}},\ }\bibfield  {title} {\bibinfo {title} {Skyrmion
  lattice in a chiral magnet},\ }\href
  {https://doi.org/10.1126/science.1166767} {\bibfield  {journal} {\bibinfo
  {journal} {Science}\ }\textbf {\bibinfo {volume} {323}},\ \bibinfo {pages}
  {915} (\bibinfo {year} {2009})}\BibitemShut {NoStop}%
\bibitem [{\citenamefont {Jaccarino}\ \emph {et~al.}(1967)\citenamefont
  {Jaccarino}, \citenamefont {Wertheim}, \citenamefont {Wernick}, \citenamefont
  {Walker},\ and\ \citenamefont {Arajs}}]{Jaccarino1967}%
  \BibitemOpen
  \bibfield  {author} {\bibinfo {author} {\bibfnamefont {V.}~\bibnamefont
  {Jaccarino}}, \bibinfo {author} {\bibfnamefont {G.~K.}\ \bibnamefont
  {Wertheim}}, \bibinfo {author} {\bibfnamefont {J.~H.}\ \bibnamefont
  {Wernick}}, \bibinfo {author} {\bibfnamefont {L.~R.}\ \bibnamefont
  {Walker}},\ and\ \bibinfo {author} {\bibfnamefont {S.}~\bibnamefont
  {Arajs}},\ }\bibfield  {title} {\bibinfo {title} {Paramagnetic excited state
  of {FeSi}},\ }\href {https://doi.org/10.1103/PhysRev.160.476} {\bibfield
  {journal} {\bibinfo  {journal} {Phys. Rev.}\ }\textbf {\bibinfo {volume}
  {160}},\ \bibinfo {pages} {476} (\bibinfo {year} {1967})}\BibitemShut
  {NoStop}%
\bibitem [{\citenamefont {Wertheim}\ \emph {et~al.}(1965)\citenamefont
  {Wertheim}, \citenamefont {Jaccarino}, \citenamefont {Wernick}, \citenamefont
  {Seitchik}, \citenamefont {Williams},\ and\ \citenamefont
  {Sherwood}}]{Wertheim1965}%
  \BibitemOpen
  \bibfield  {author} {\bibinfo {author} {\bibfnamefont {G.}~\bibnamefont
  {Wertheim}}, \bibinfo {author} {\bibfnamefont {V.}~\bibnamefont {Jaccarino}},
  \bibinfo {author} {\bibfnamefont {J.}~\bibnamefont {Wernick}}, \bibinfo
  {author} {\bibfnamefont {J.}~\bibnamefont {Seitchik}}, \bibinfo {author}
  {\bibfnamefont {H.}~\bibnamefont {Williams}},\ and\ \bibinfo {author}
  {\bibfnamefont {R.}~\bibnamefont {Sherwood}},\ }\bibfield  {title} {\bibinfo
  {title} {Unusual electronic properties of {FeSi}},\ }\href
  {https://doi.org/https://doi.org/10.1016/0031-9163(65)90658-X} {\bibfield
  {journal} {\bibinfo  {journal} {Physics Letters}\ }\textbf {\bibinfo {volume}
  {18}},\ \bibinfo {pages} {89} (\bibinfo {year} {1965})}\BibitemShut {NoStop}%
\bibitem [{\citenamefont {Mandrus}\ \emph {et~al.}(1995)\citenamefont
  {Mandrus}, \citenamefont {Sarrao}, \citenamefont {Migliori}, \citenamefont
  {Thompson},\ and\ \citenamefont {Fisk}}]{Mandrus1995}%
  \BibitemOpen
  \bibfield  {author} {\bibinfo {author} {\bibfnamefont {D.}~\bibnamefont
  {Mandrus}}, \bibinfo {author} {\bibfnamefont {J.~L.}\ \bibnamefont {Sarrao}},
  \bibinfo {author} {\bibfnamefont {A.}~\bibnamefont {Migliori}}, \bibinfo
  {author} {\bibfnamefont {J.~D.}\ \bibnamefont {Thompson}},\ and\ \bibinfo
  {author} {\bibfnamefont {Z.}~\bibnamefont {Fisk}},\ }\bibfield  {title}
  {\bibinfo {title} {Thermodynamics of {FeSi}},\ }\href
  {https://doi.org/10.1103/PhysRevB.51.4763} {\bibfield  {journal} {\bibinfo
  {journal} {Phys. Rev. B}\ }\textbf {\bibinfo {volume} {51}},\ \bibinfo
  {pages} {4763} (\bibinfo {year} {1995})}\BibitemShut {NoStop}%
\bibitem [{\citenamefont {Schlesinger}\ \emph {et~al.}(1993)\citenamefont
  {Schlesinger}, \citenamefont {Fisk}, \citenamefont {Zhang}, \citenamefont
  {Maple}, \citenamefont {DiTusa},\ and\ \citenamefont
  {Aeppli}}]{Schlesinger1993}%
  \BibitemOpen
  \bibfield  {author} {\bibinfo {author} {\bibfnamefont {Z.}~\bibnamefont
  {Schlesinger}}, \bibinfo {author} {\bibfnamefont {Z.}~\bibnamefont {Fisk}},
  \bibinfo {author} {\bibfnamefont {H.-T.}\ \bibnamefont {Zhang}}, \bibinfo
  {author} {\bibfnamefont {M.~B.}\ \bibnamefont {Maple}}, \bibinfo {author}
  {\bibfnamefont {J.}~\bibnamefont {DiTusa}},\ and\ \bibinfo {author}
  {\bibfnamefont {G.}~\bibnamefont {Aeppli}},\ }\bibfield  {title} {\bibinfo
  {title} {Unconventional charge gap formation in {FeSi}},\ }\href
  {https://doi.org/10.1103/PhysRevLett.71.1748} {\bibfield  {journal} {\bibinfo
   {journal} {Phys. Rev. Lett.}\ }\textbf {\bibinfo {volume} {71}},\ \bibinfo
  {pages} {1748} (\bibinfo {year} {1993})}\BibitemShut {NoStop}%
\bibitem [{\citenamefont {Degiorgi}\ \emph {et~al.}(1994)\citenamefont
  {Degiorgi}, \citenamefont {Hunt}, \citenamefont {Ott}, \citenamefont
  {Dressel}, \citenamefont {Feenstra}, \citenamefont {Gr\"{u}uner},
  \citenamefont {Fisk},\ and\ \citenamefont {Canfield}}]{Degiorgi1994}%
  \BibitemOpen
  \bibfield  {author} {\bibinfo {author} {\bibfnamefont {L.}~\bibnamefont
  {Degiorgi}}, \bibinfo {author} {\bibfnamefont {M.~B.}\ \bibnamefont {Hunt}},
  \bibinfo {author} {\bibfnamefont {H.~R.}\ \bibnamefont {Ott}}, \bibinfo
  {author} {\bibfnamefont {M.}~\bibnamefont {Dressel}}, \bibinfo {author}
  {\bibfnamefont {B.~J.}\ \bibnamefont {Feenstra}}, \bibinfo {author}
  {\bibfnamefont {G.}~\bibnamefont {Gr\"{u}uner}}, \bibinfo {author}
  {\bibfnamefont {Z.}~\bibnamefont {Fisk}},\ and\ \bibinfo {author}
  {\bibfnamefont {P.}~\bibnamefont {Canfield}},\ }\bibfield  {title} {\bibinfo
  {title} {Optical evidence of anderson-mott localization in {FeSi}},\ }\href
  {https://doi.org/10.1209/0295-5075/28/5/008} {\bibfield  {journal} {\bibinfo
  {journal} {Europhysics Letters ({EPL})}\ }\textbf {\bibinfo {volume} {28}},\
  \bibinfo {pages} {341} (\bibinfo {year} {1994})}\BibitemShut {NoStop}%
\bibitem [{\citenamefont {Wolfe}\ \emph {et~al.}(1965)\citenamefont {Wolfe},
  \citenamefont {Wernick},\ and\ \citenamefont {Haszko}}]{Wolfe1965}%
  \BibitemOpen
  \bibfield  {author} {\bibinfo {author} {\bibfnamefont {R.}~\bibnamefont
  {Wolfe}}, \bibinfo {author} {\bibfnamefont {J.}~\bibnamefont {Wernick}},\
  and\ \bibinfo {author} {\bibfnamefont {S.}~\bibnamefont {Haszko}},\
  }\bibfield  {title} {\bibinfo {title} {Thermoelectric properties of {FeSi}},\
  }\href {https://doi.org/https://doi.org/10.1016/0031-9163(65)90094-6}
  {\bibfield  {journal} {\bibinfo  {journal} {Physics Letters}\ }\textbf
  {\bibinfo {volume} {19}},\ \bibinfo {pages} {449} (\bibinfo {year}
  {1965})}\BibitemShut {NoStop}%
\bibitem [{\citenamefont {Buschinger}\ \emph {et~al.}(1997)\citenamefont
  {Buschinger}, \citenamefont {Geibel}, \citenamefont {Steglich}, \citenamefont
  {Mandrus}, \citenamefont {Young}, \citenamefont {Sarrao},\ and\ \citenamefont
  {Fisk}}]{Buschinger1997}%
  \BibitemOpen
  \bibfield  {author} {\bibinfo {author} {\bibfnamefont {B.}~\bibnamefont
  {Buschinger}}, \bibinfo {author} {\bibfnamefont {C.}~\bibnamefont {Geibel}},
  \bibinfo {author} {\bibfnamefont {F.}~\bibnamefont {Steglich}}, \bibinfo
  {author} {\bibfnamefont {D.}~\bibnamefont {Mandrus}}, \bibinfo {author}
  {\bibfnamefont {D.}~\bibnamefont {Young}}, \bibinfo {author} {\bibfnamefont
  {J.}~\bibnamefont {Sarrao}},\ and\ \bibinfo {author} {\bibfnamefont
  {Z.}~\bibnamefont {Fisk}},\ }\bibfield  {title} {\bibinfo {title} {Transport
  properties of {FeSi}},\ }\href
  {https://doi.org/https://doi.org/10.1016/S0921-4526(96)00839-3} {\bibfield
  {journal} {\bibinfo  {journal} {Physica B: Condensed Matter}\ }\textbf
  {\bibinfo {volume} {230-232}},\ \bibinfo {pages} {784} (\bibinfo {year}
  {1997})},\ \bibinfo {note} {proceedings of the International Conference on
  Strongly Correlated Electron Systems}\BibitemShut {NoStop}%
\bibitem [{\citenamefont {Paschen}\ \emph {et~al.}(1997)\citenamefont
  {Paschen}, \citenamefont {Felder}, \citenamefont {Chernikov}, \citenamefont
  {Degiorgi}, \citenamefont {Schwer}, \citenamefont {Ott}, \citenamefont
  {Young}, \citenamefont {Sarrao},\ and\ \citenamefont {Fisk}}]{Paschen1997}%
  \BibitemOpen
  \bibfield  {author} {\bibinfo {author} {\bibfnamefont {S.}~\bibnamefont
  {Paschen}}, \bibinfo {author} {\bibfnamefont {E.}~\bibnamefont {Felder}},
  \bibinfo {author} {\bibfnamefont {M.~A.}\ \bibnamefont {Chernikov}}, \bibinfo
  {author} {\bibfnamefont {L.}~\bibnamefont {Degiorgi}}, \bibinfo {author}
  {\bibfnamefont {H.}~\bibnamefont {Schwer}}, \bibinfo {author} {\bibfnamefont
  {H.~R.}\ \bibnamefont {Ott}}, \bibinfo {author} {\bibfnamefont {D.~P.}\
  \bibnamefont {Young}}, \bibinfo {author} {\bibfnamefont {J.~L.}\ \bibnamefont
  {Sarrao}},\ and\ \bibinfo {author} {\bibfnamefont {Z.}~\bibnamefont {Fisk}},\
  }\bibfield  {title} {\bibinfo {title} {Low-temperature transport,
  thermodynamic, and optical properties of {FeSi}},\ }\href
  {https://doi.org/10.1103/PhysRevB.56.12916} {\bibfield  {journal} {\bibinfo
  {journal} {Phys. Rev. B}\ }\textbf {\bibinfo {volume} {56}},\ \bibinfo
  {pages} {12916} (\bibinfo {year} {1997})}\BibitemShut {NoStop}%
\bibitem [{\citenamefont {{van der Marel}}\ \emph {et~al.}(1998)\citenamefont
  {{van der Marel}}, \citenamefont {Damascelli}, \citenamefont {Schulte},\ and\
  \citenamefont {Menovsky}}]{vanderMarel1998}%
  \BibitemOpen
  \bibfield  {author} {\bibinfo {author} {\bibfnamefont {D.}~\bibnamefont {{van
  der Marel}}}, \bibinfo {author} {\bibfnamefont {A.}~\bibnamefont
  {Damascelli}}, \bibinfo {author} {\bibfnamefont {K.}~\bibnamefont
  {Schulte}},\ and\ \bibinfo {author} {\bibfnamefont {A.}~\bibnamefont
  {Menovsky}},\ }\bibfield  {title} {\bibinfo {title} {Spin, charge, and
  bonding in transition metal mono-silicides},\ }\href
  {https://doi.org/https://doi.org/10.1016/S0921-4526(97)00476-6} {\bibfield
  {journal} {\bibinfo  {journal} {Physica B: Condensed Matter}\ }\textbf
  {\bibinfo {volume} {244}},\ \bibinfo {pages} {138} (\bibinfo {year}
  {1998})}\BibitemShut {NoStop}%
\bibitem [{\citenamefont {Damascelli}\ \emph {et~al.}(1997)\citenamefont
  {Damascelli}, \citenamefont {Schulte}, \citenamefont {van~der Marel},\ and\
  \citenamefont {Menovsky}}]{Damascelli1997}%
  \BibitemOpen
  \bibfield  {author} {\bibinfo {author} {\bibfnamefont {A.}~\bibnamefont
  {Damascelli}}, \bibinfo {author} {\bibfnamefont {K.}~\bibnamefont {Schulte}},
  \bibinfo {author} {\bibfnamefont {D.}~\bibnamefont {van~der Marel}},\ and\
  \bibinfo {author} {\bibfnamefont {A.~A.}\ \bibnamefont {Menovsky}},\
  }\bibfield  {title} {\bibinfo {title} {Infrared spectroscopic study of
  phonons coupled to charge excitations in {FeSi}},\ }\href
  {https://doi.org/10.1103/PhysRevB.55.R4863} {\bibfield  {journal} {\bibinfo
  {journal} {Phys. Rev. B}\ }\textbf {\bibinfo {volume} {55}},\ \bibinfo
  {pages} {R4863} (\bibinfo {year} {1997})}\BibitemShut {NoStop}%
\bibitem [{\citenamefont {Menzel}\ \emph {et~al.}(2009)\citenamefont {Menzel},
  \citenamefont {Popovich}, \citenamefont {Kovaleva}, \citenamefont {Schoenes},
  \citenamefont {Doll},\ and\ \citenamefont {Boris}}]{Menzel2009}%
  \BibitemOpen
  \bibfield  {author} {\bibinfo {author} {\bibfnamefont {D.}~\bibnamefont
  {Menzel}}, \bibinfo {author} {\bibfnamefont {P.}~\bibnamefont {Popovich}},
  \bibinfo {author} {\bibfnamefont {N.~N.}\ \bibnamefont {Kovaleva}}, \bibinfo
  {author} {\bibfnamefont {J.}~\bibnamefont {Schoenes}}, \bibinfo {author}
  {\bibfnamefont {K.}~\bibnamefont {Doll}},\ and\ \bibinfo {author}
  {\bibfnamefont {A.~V.}\ \bibnamefont {Boris}},\ }\bibfield  {title} {\bibinfo
  {title} {Electron-phonon interaction and spectral weight transfer in
  {${\text{Fe}}_{1\ensuremath{-}x}{\text{Co}}_{x}\text{Si}$}},\ }\href
  {https://doi.org/10.1103/PhysRevB.79.165111} {\bibfield  {journal} {\bibinfo
  {journal} {Phys. Rev. B}\ }\textbf {\bibinfo {volume} {79}},\ \bibinfo
  {pages} {165111} (\bibinfo {year} {2009})}\BibitemShut {NoStop}%
\bibitem [{\citenamefont {Fang}\ \emph {et~al.}(2018)\citenamefont {Fang},
  \citenamefont {Ran}, \citenamefont {Xie}, \citenamefont {Wang}, \citenamefont
  {Meng},\ and\ \citenamefont {Maple}}]{Fang2018}%
  \BibitemOpen
  \bibfield  {author} {\bibinfo {author} {\bibfnamefont {Y.}~\bibnamefont
  {Fang}}, \bibinfo {author} {\bibfnamefont {S.}~\bibnamefont {Ran}}, \bibinfo
  {author} {\bibfnamefont {W.}~\bibnamefont {Xie}}, \bibinfo {author}
  {\bibfnamefont {S.}~\bibnamefont {Wang}}, \bibinfo {author} {\bibfnamefont
  {Y.~S.}\ \bibnamefont {Meng}},\ and\ \bibinfo {author} {\bibfnamefont
  {M.~B.}\ \bibnamefont {Maple}},\ }\bibfield  {title} {\bibinfo {title}
  {Evidence for a conducting surface ground state in high-quality single
  crystalline fesi},\ }\href {https://doi.org/10.1073/pnas.1806910115}
  {\bibfield  {journal} {\bibinfo  {journal} {Proceedings of the National
  Academy of Sciences}\ }\textbf {\bibinfo {volume} {115}},\ \bibinfo {pages}
  {8558} (\bibinfo {year} {2018})}\BibitemShut {NoStop}%
\bibitem [{\citenamefont {Takagi}\ \emph {et~al.}(1981)\citenamefont {Takagi},
  \citenamefont {Yasuoka}, \citenamefont {Ogawa},\ and\ \citenamefont
  {H.~Wernick}}]{Takagi1981}%
  \BibitemOpen
  \bibfield  {author} {\bibinfo {author} {\bibfnamefont {S.}~\bibnamefont
  {Takagi}}, \bibinfo {author} {\bibfnamefont {H.}~\bibnamefont {Yasuoka}},
  \bibinfo {author} {\bibfnamefont {S.}~\bibnamefont {Ogawa}},\ and\ \bibinfo
  {author} {\bibfnamefont {J.}~\bibnamefont {H.~Wernick}},\ }\bibfield  {title}
  {\bibinfo {title} {{{}$^{29}$Si NMR} studies of an ``unusual" paramagnet
  {FeSi}$-${Anderson Localized State Model}$-$},\ }\href
  {https://doi.org/10.1143/JPSJ.50.2539} {\bibfield  {journal} {\bibinfo
  {journal} {Journal of the Physical Society of Japan}\ }\textbf {\bibinfo
  {volume} {50}},\ \bibinfo {pages} {2539} (\bibinfo {year}
  {1981})}\BibitemShut {NoStop}%
\bibitem [{\citenamefont {Delaire}\ \emph {et~al.}(2011)\citenamefont
  {Delaire}, \citenamefont {Marty}, \citenamefont {Stone}, \citenamefont
  {Kent}, \citenamefont {Lucas}, \citenamefont {Abernathy}, \citenamefont
  {Mandrus},\ and\ \citenamefont {Sales}}]{Delaire2011}%
  \BibitemOpen
  \bibfield  {author} {\bibinfo {author} {\bibfnamefont {O.}~\bibnamefont
  {Delaire}}, \bibinfo {author} {\bibfnamefont {K.}~\bibnamefont {Marty}},
  \bibinfo {author} {\bibfnamefont {M.~B.}\ \bibnamefont {Stone}}, \bibinfo
  {author} {\bibfnamefont {P.~R.~C.}\ \bibnamefont {Kent}}, \bibinfo {author}
  {\bibfnamefont {M.~S.}\ \bibnamefont {Lucas}}, \bibinfo {author}
  {\bibfnamefont {D.~L.}\ \bibnamefont {Abernathy}}, \bibinfo {author}
  {\bibfnamefont {D.}~\bibnamefont {Mandrus}},\ and\ \bibinfo {author}
  {\bibfnamefont {B.~C.}\ \bibnamefont {Sales}},\ }\bibfield  {title} {\bibinfo
  {title} {Phonon softening and metallization of a narrow-gap semiconductor by
  thermal disorder},\ }\href {https://doi.org/10.1073/pnas.1014869108}
  {\bibfield  {journal} {\bibinfo  {journal} {Proceedings of the National
  Academy of Sciences}\ }\textbf {\bibinfo {volume} {108}},\ \bibinfo {pages}
  {4725} (\bibinfo {year} {2011})}\BibitemShut {NoStop}%
\bibitem [{\citenamefont {Krannich}\ \emph {et~al.}(2015)\citenamefont
  {Krannich}, \citenamefont {Sidis}, \citenamefont {Lamago}, \citenamefont
  {Heid}, \citenamefont {Mignot}, \citenamefont {v.~L\"ohneysen}, \citenamefont
  {Ivanov}, \citenamefont {Steffens}, \citenamefont {Keller}, \citenamefont
  {Wang}, \citenamefont {Goering},\ and\ \citenamefont {Weber}}]{Krannich2015}%
  \BibitemOpen
  \bibfield  {author} {\bibinfo {author} {\bibfnamefont {S.}~\bibnamefont
  {Krannich}}, \bibinfo {author} {\bibfnamefont {Y.}~\bibnamefont {Sidis}},
  \bibinfo {author} {\bibfnamefont {D.}~\bibnamefont {Lamago}}, \bibinfo
  {author} {\bibfnamefont {R.}~\bibnamefont {Heid}}, \bibinfo {author}
  {\bibfnamefont {J.-M.}\ \bibnamefont {Mignot}}, \bibinfo {author}
  {\bibfnamefont {H.}~\bibnamefont {v.~L\"ohneysen}}, \bibinfo {author}
  {\bibfnamefont {A.}~\bibnamefont {Ivanov}}, \bibinfo {author} {\bibfnamefont
  {P.}~\bibnamefont {Steffens}}, \bibinfo {author} {\bibfnamefont
  {T.}~\bibnamefont {Keller}}, \bibinfo {author} {\bibfnamefont
  {L.}~\bibnamefont {Wang}}, \bibinfo {author} {\bibfnamefont {E.}~\bibnamefont
  {Goering}},\ and\ \bibinfo {author} {\bibfnamefont {F.}~\bibnamefont
  {Weber}},\ }\bibfield  {title} {\bibinfo {title} {Magnetic moments induce
  strong phonon renormalization in {FeSi}},\ }\href
  {https://doi.org/10.1038/ncomms9961} {\bibfield  {journal} {\bibinfo
  {journal} {Nature Communications}\ }\textbf {\bibinfo {volume} {6}},\
  \bibinfo {pages} {8961} (\bibinfo {year} {2015})}\BibitemShut {NoStop}%
\bibitem [{\citenamefont {Tomczak}\ \emph {et~al.}(2012)\citenamefont
  {Tomczak}, \citenamefont {Haule},\ and\ \citenamefont
  {Kotliar}}]{Tomczak2012}%
  \BibitemOpen
  \bibfield  {author} {\bibinfo {author} {\bibfnamefont {J.~M.}\ \bibnamefont
  {Tomczak}}, \bibinfo {author} {\bibfnamefont {K.}~\bibnamefont {Haule}},\
  and\ \bibinfo {author} {\bibfnamefont {G.}~\bibnamefont {Kotliar}},\
  }\bibfield  {title} {\bibinfo {title} {Signatures of electronic correlations
  in iron silicide},\ }\href {https://doi.org/10.1073/pnas.1118371109}
  {\bibfield  {journal} {\bibinfo  {journal} {Proceedings of the National
  Academy of Sciences}\ }\textbf {\bibinfo {volume} {109}},\ \bibinfo {pages}
  {3243} (\bibinfo {year} {2012})}\BibitemShut {NoStop}%
\bibitem [{\citenamefont {Jarlborg}(1999)}]{Jarlborg1999}%
  \BibitemOpen
  \bibfield  {author} {\bibinfo {author} {\bibfnamefont {T.}~\bibnamefont
  {Jarlborg}},\ }\bibfield  {title} {\bibinfo {title} {Electronic structure and
  properties of pure and doped {$\ensuremath{\epsilon}$-FeSi} from ab initio
  local-density theory},\ }\href {https://doi.org/10.1103/PhysRevB.59.15002}
  {\bibfield  {journal} {\bibinfo  {journal} {Phys. Rev. B}\ }\textbf {\bibinfo
  {volume} {59}},\ \bibinfo {pages} {15002} (\bibinfo {year}
  {1999})}\BibitemShut {NoStop}%
\bibitem [{\citenamefont {Mazurenko}\ \emph {et~al.}(2010)\citenamefont
  {Mazurenko}, \citenamefont {Shorikov}, \citenamefont {Lukoyanov},
  \citenamefont {Kharlov}, \citenamefont {Gorelov}, \citenamefont
  {Lichtenstein},\ and\ \citenamefont {Anisimov}}]{Mazurenko2010}%
  \BibitemOpen
  \bibfield  {author} {\bibinfo {author} {\bibfnamefont {V.~V.}\ \bibnamefont
  {Mazurenko}}, \bibinfo {author} {\bibfnamefont {A.~O.}\ \bibnamefont
  {Shorikov}}, \bibinfo {author} {\bibfnamefont {A.~V.}\ \bibnamefont
  {Lukoyanov}}, \bibinfo {author} {\bibfnamefont {K.}~\bibnamefont {Kharlov}},
  \bibinfo {author} {\bibfnamefont {E.}~\bibnamefont {Gorelov}}, \bibinfo
  {author} {\bibfnamefont {A.~I.}\ \bibnamefont {Lichtenstein}},\ and\ \bibinfo
  {author} {\bibfnamefont {V.~I.}\ \bibnamefont {Anisimov}},\ }\bibfield
  {title} {\bibinfo {title} {Metal-insulator transitions and magnetism in
  correlated band insulators: {FeSi} and
  {${\text{Fe}}_{1\ensuremath{-}x}{\text{Co}}_{x}\text{Si}$}},\ }\href
  {https://doi.org/10.1103/PhysRevB.81.125131} {\bibfield  {journal} {\bibinfo
  {journal} {Phys. Rev. B}\ }\textbf {\bibinfo {volume} {81}},\ \bibinfo
  {pages} {125131} (\bibinfo {year} {2010})}\BibitemShut {NoStop}%
\bibitem [{\citenamefont {Kune\ifmmode~\check{s}\else \v{s}\fi{}}\ and\
  \citenamefont {Anisimov}(2008)}]{Kune2008}%
  \BibitemOpen
  \bibfield  {author} {\bibinfo {author} {\bibfnamefont {J.}~\bibnamefont
  {Kune\ifmmode~\check{s}\else \v{s}\fi{}}}\ and\ \bibinfo {author}
  {\bibfnamefont {V.~I.}\ \bibnamefont {Anisimov}},\ }\bibfield  {title}
  {\bibinfo {title} {Temperature-dependent correlations in covalent insulators:
  Dynamical mean-field approximation},\ }\href
  {https://doi.org/10.1103/PhysRevB.78.033109} {\bibfield  {journal} {\bibinfo
  {journal} {Phys. Rev. B}\ }\textbf {\bibinfo {volume} {78}},\ \bibinfo
  {pages} {033109} (\bibinfo {year} {2008})}\BibitemShut {NoStop}%
\bibitem [{\citenamefont {Heid}\ and\ \citenamefont {Bohnen}(1999)}]{Heid1999}%
  \BibitemOpen
  \bibfield  {author} {\bibinfo {author} {\bibfnamefont {R.}~\bibnamefont
  {Heid}}\ and\ \bibinfo {author} {\bibfnamefont {K.-P.}\ \bibnamefont
  {Bohnen}},\ }\bibfield  {title} {\bibinfo {title} {Linear response in a
  density-functional mixed-basis approach},\ }\href
  {https://doi.org/10.1103/PhysRevB.60.R3709} {\bibfield  {journal} {\bibinfo
  {journal} {Phys. Rev. B}\ }\textbf {\bibinfo {volume} {60}},\ \bibinfo
  {pages} {R3709} (\bibinfo {year} {1999})}\BibitemShut {NoStop}%
\bibitem [{\citenamefont {Perdew}\ and\ \citenamefont
  {Wang}(1992)}]{Perdew1992}%
  \BibitemOpen
  \bibfield  {author} {\bibinfo {author} {\bibfnamefont {J.~P.}\ \bibnamefont
  {Perdew}}\ and\ \bibinfo {author} {\bibfnamefont {Y.}~\bibnamefont {Wang}},\
  }\bibfield  {title} {\bibinfo {title} {Accurate and simple analytic
  representation of the electron-gas correlation energy},\ }\href
  {https://doi.org/10.1103/PhysRevB.45.13244} {\bibfield  {journal} {\bibinfo
  {journal} {Phys. Rev. B}\ }\textbf {\bibinfo {volume} {45}},\ \bibinfo
  {pages} {13244} (\bibinfo {year} {1992})}\BibitemShut {NoStop}%
\bibitem [{\citenamefont {Vanderbilt}(1985)}]{Vanderbilt1985}%
  \BibitemOpen
  \bibfield  {author} {\bibinfo {author} {\bibfnamefont {D.}~\bibnamefont
  {Vanderbilt}},\ }\bibfield  {title} {\bibinfo {title} {Optimally smooth
  norm-conserving pseudopotentials},\ }\href
  {https://doi.org/10.1103/PhysRevB.32.8412} {\bibfield  {journal} {\bibinfo
  {journal} {Phys. Rev. B}\ }\textbf {\bibinfo {volume} {32}},\ \bibinfo
  {pages} {8412} (\bibinfo {year} {1985})}\BibitemShut {NoStop}%
\bibitem [{\citenamefont {Bewley}\ \emph {et~al.}(2006)\citenamefont {Bewley},
  \citenamefont {Eccleston}, \citenamefont {McEwen}, \citenamefont {Hayden},
  \citenamefont {Dove}, \citenamefont {Bennington}, \citenamefont {Treadgold},\
  and\ \citenamefont {Coleman}}]{bewley2006}%
  \BibitemOpen
  \bibfield  {author} {\bibinfo {author} {\bibfnamefont {R.}~\bibnamefont
  {Bewley}}, \bibinfo {author} {\bibfnamefont {R.}~\bibnamefont {Eccleston}},
  \bibinfo {author} {\bibfnamefont {K.}~\bibnamefont {McEwen}}, \bibinfo
  {author} {\bibfnamefont {S.}~\bibnamefont {Hayden}}, \bibinfo {author}
  {\bibfnamefont {M.}~\bibnamefont {Dove}}, \bibinfo {author} {\bibfnamefont
  {S.}~\bibnamefont {Bennington}}, \bibinfo {author} {\bibfnamefont
  {J.}~\bibnamefont {Treadgold}},\ and\ \bibinfo {author} {\bibfnamefont
  {R.}~\bibnamefont {Coleman}},\ }\bibfield  {title} {\bibinfo {title} {Merlin,
  a new high count rate spectrometer at isis},\ }\href
  {https://doi.org/https://doi.org/10.1016/j.physb.2006.05.328} {\bibfield
  {journal} {\bibinfo  {journal} {Physica B: Condensed Matter}\ }\textbf
  {\bibinfo {volume} {385-386}},\ \bibinfo {pages} {1029} (\bibinfo {year}
  {2006})}\BibitemShut {NoStop}%
\bibitem [{\citenamefont {Weber}\ \emph {et~al.}(2020)\citenamefont {Weber},
  \citenamefont {Khan}, \citenamefont {Walker},\ and\ \citenamefont
  {Voneshen}}]{isisdata2020}%
  \BibitemOpen
  \bibfield  {author} {\bibinfo {author} {\bibfnamefont {F.}~\bibnamefont
  {Weber}}, \bibinfo {author} {\bibfnamefont {N.}~\bibnamefont {Khan}},
  \bibinfo {author} {\bibfnamefont {H.}~\bibnamefont {Walker}},\ and\ \bibinfo
  {author} {\bibfnamefont {D.}~\bibnamefont {Voneshen}},\ }\bibfield  {title}
  {\bibinfo {title} {Spin-phonon coupling in {Mn$_{1-x}$Fe$_{x}$Si}},\
  }\bibfield  {journal} {\bibinfo  {journal} {STFC ISIS Neutron and Muon
  Source}\ }\href {https://doi.org/10.5286/ISIS.E.RB2010794}
  {10.5286/ISIS.E.RB2010794} (\bibinfo {year} {2020})\BibitemShut {NoStop}%
\bibitem [{\citenamefont {Arnold}\ \emph {et~al.}(2014)\citenamefont {Arnold},
  \citenamefont {Bilheux}, \citenamefont {Borreguero}, \citenamefont {Buts},
  \citenamefont {Campbell}, \citenamefont {Chapon}, \citenamefont {Doucet},
  \citenamefont {Draper}, \citenamefont {{Ferraz Leal}}, \citenamefont {Gigg},
  \citenamefont {Lynch}, \citenamefont {Markvardsen}, \citenamefont
  {Mikkelson}, \citenamefont {Mikkelson}, \citenamefont {Miller}, \citenamefont
  {Palmen}, \citenamefont {Parker}, \citenamefont {Passos}, \citenamefont
  {Perring}, \citenamefont {Peterson}, \citenamefont {Ren}, \citenamefont
  {Reuter}, \citenamefont {Savici}, \citenamefont {Taylor}, \citenamefont
  {Taylor}, \citenamefont {Tolchenov}, \citenamefont {Zhou},\ and\
  \citenamefont {Zikovsky}}]{ARNOLD2014}%
  \BibitemOpen
  \bibfield  {author} {\bibinfo {author} {\bibfnamefont {O.}~\bibnamefont
  {Arnold}}, \bibinfo {author} {\bibfnamefont {J.}~\bibnamefont {Bilheux}},
  \bibinfo {author} {\bibfnamefont {J.}~\bibnamefont {Borreguero}}, \bibinfo
  {author} {\bibfnamefont {A.}~\bibnamefont {Buts}}, \bibinfo {author}
  {\bibfnamefont {S.}~\bibnamefont {Campbell}}, \bibinfo {author}
  {\bibfnamefont {L.}~\bibnamefont {Chapon}}, \bibinfo {author} {\bibfnamefont
  {M.}~\bibnamefont {Doucet}}, \bibinfo {author} {\bibfnamefont
  {N.}~\bibnamefont {Draper}}, \bibinfo {author} {\bibfnamefont
  {R.}~\bibnamefont {{Ferraz Leal}}}, \bibinfo {author} {\bibfnamefont
  {M.}~\bibnamefont {Gigg}}, \bibinfo {author} {\bibfnamefont {V.}~\bibnamefont
  {Lynch}}, \bibinfo {author} {\bibfnamefont {A.}~\bibnamefont {Markvardsen}},
  \bibinfo {author} {\bibfnamefont {D.}~\bibnamefont {Mikkelson}}, \bibinfo
  {author} {\bibfnamefont {R.}~\bibnamefont {Mikkelson}}, \bibinfo {author}
  {\bibfnamefont {R.}~\bibnamefont {Miller}}, \bibinfo {author} {\bibfnamefont
  {K.}~\bibnamefont {Palmen}}, \bibinfo {author} {\bibfnamefont
  {P.}~\bibnamefont {Parker}}, \bibinfo {author} {\bibfnamefont
  {G.}~\bibnamefont {Passos}}, \bibinfo {author} {\bibfnamefont
  {T.}~\bibnamefont {Perring}}, \bibinfo {author} {\bibfnamefont
  {P.}~\bibnamefont {Peterson}}, \bibinfo {author} {\bibfnamefont
  {S.}~\bibnamefont {Ren}}, \bibinfo {author} {\bibfnamefont {M.}~\bibnamefont
  {Reuter}}, \bibinfo {author} {\bibfnamefont {A.}~\bibnamefont {Savici}},
  \bibinfo {author} {\bibfnamefont {J.}~\bibnamefont {Taylor}}, \bibinfo
  {author} {\bibfnamefont {R.}~\bibnamefont {Taylor}}, \bibinfo {author}
  {\bibfnamefont {R.}~\bibnamefont {Tolchenov}}, \bibinfo {author}
  {\bibfnamefont {W.}~\bibnamefont {Zhou}},\ and\ \bibinfo {author}
  {\bibfnamefont {J.}~\bibnamefont {Zikovsky}},\ }\bibfield  {title} {\bibinfo
  {title} {Mantid$-${Data} analysis and visualization package for neutron
  scattering and {$\mu$SR} experiments},\ }\href
  {https://doi.org/https://doi.org/10.1016/j.nima.2014.07.029} {\bibfield
  {journal} {\bibinfo  {journal} {Nuclear Instruments and Methods in Physics
  Research Section A: Accelerators, Spectrometers, Detectors and Associated
  Equipment}\ }\textbf {\bibinfo {volume} {764}},\ \bibinfo {pages} {156}
  (\bibinfo {year} {2014})}\BibitemShut {NoStop}%
\bibitem [{\citenamefont {Ewings}\ \emph {et~al.}(2016)\citenamefont {Ewings},
  \citenamefont {Buts}, \citenamefont {Le}, \citenamefont {{van Duijn}},
  \citenamefont {Bustinduy},\ and\ \citenamefont {Perring}}]{EWINGS2016}%
  \BibitemOpen
  \bibfield  {author} {\bibinfo {author} {\bibfnamefont {R.}~\bibnamefont
  {Ewings}}, \bibinfo {author} {\bibfnamefont {A.}~\bibnamefont {Buts}},
  \bibinfo {author} {\bibfnamefont {M.}~\bibnamefont {Le}}, \bibinfo {author}
  {\bibfnamefont {J.}~\bibnamefont {{van Duijn}}}, \bibinfo {author}
  {\bibfnamefont {I.}~\bibnamefont {Bustinduy}},\ and\ \bibinfo {author}
  {\bibfnamefont {T.}~\bibnamefont {Perring}},\ }\bibfield  {title} {\bibinfo
  {title} {Horace: Software for the analysis of data from single crystal
  spectroscopy experiments at time-of-flight neutron instruments},\ }\href
  {https://doi.org/https://doi.org/10.1016/j.nima.2016.07.036} {\bibfield
  {journal} {\bibinfo  {journal} {Nuclear Instruments and Methods in Physics
  Research Section A: Accelerators, Spectrometers, Detectors and Associated
  Equipment}\ }\textbf {\bibinfo {volume} {834}},\ \bibinfo {pages} {132}
  (\bibinfo {year} {2016})}\BibitemShut {NoStop}%
\bibitem [{\citenamefont {Moertter}\ \emph {et~al.}(2016)\citenamefont
  {Moertter}, \citenamefont {Boll}, \citenamefont {Ivanov}, \citenamefont
  {Sidis},\ and\ \citenamefont {Weber}}]{illdata2016}%
  \BibitemOpen
  \bibfield  {author} {\bibinfo {author} {\bibfnamefont {M.}~\bibnamefont
  {Moertter}}, \bibinfo {author} {\bibfnamefont {D.}~\bibnamefont {Boll}},
  \bibinfo {author} {\bibfnamefont {A.}~\bibnamefont {Ivanov}}, \bibinfo
  {author} {\bibfnamefont {Y.}~\bibnamefont {Sidis}},\ and\ \bibinfo {author}
  {\bibfnamefont {F.}~\bibnamefont {Weber}},\ }\bibfield  {title} {\bibinfo
  {title} {Spin-phonon coupling in {FeSi}},\ }\bibfield  {journal} {\bibinfo
  {journal} {Institut Laue-Langevin}\ }\href
  {https://doi.org/10.5291/ILL-DATA.7-01-439} {10.5291/ILL-DATA.7-01-439}
  (\bibinfo {year} {2016})\BibitemShut {NoStop}%
\bibitem [{\citenamefont {Vo\v{c}adlo}\ \emph {et~al.}(2002)\citenamefont
  {Vo\v{c}adlo}, \citenamefont {Knight}, \citenamefont {Price},\ and\
  \citenamefont {Wood}}]{Vocadlo2002}%
  \BibitemOpen
  \bibfield  {author} {\bibinfo {author} {\bibfnamefont {L.}~\bibnamefont
  {Vo\v{c}adlo}}, \bibinfo {author} {\bibfnamefont {K.~S.}\ \bibnamefont
  {Knight}}, \bibinfo {author} {\bibfnamefont {G.~D.}\ \bibnamefont {Price}},\
  and\ \bibinfo {author} {\bibfnamefont {I.~G.}\ \bibnamefont {Wood}},\
  }\bibfield  {title} {\bibinfo {title} {Thermal expansion and crystal
  structure of fesi between 4 and 1173 {K} determined by time-of-flight neutron
  powder diffraction},\ }\href {https://doi.org/10.1007/s002690100202}
  {\bibfield  {journal} {\bibinfo  {journal} {Physics and Chemistry of
  Minerals}\ }\textbf {\bibinfo {volume} {29}},\ \bibinfo {pages} {132}
  (\bibinfo {year} {2002})}\BibitemShut {NoStop}%
\bibitem [{\citenamefont {Sales}\ \emph {et~al.}(1994)\citenamefont {Sales},
  \citenamefont {Jones}, \citenamefont {Chakoumakos}, \citenamefont
  {Fernandez-Baca}, \citenamefont {Harmon}, \citenamefont {Sharp},\ and\
  \citenamefont {Volckmann}}]{Sales1994}%
  \BibitemOpen
  \bibfield  {author} {\bibinfo {author} {\bibfnamefont {B.~C.}\ \bibnamefont
  {Sales}}, \bibinfo {author} {\bibfnamefont {E.~C.}\ \bibnamefont {Jones}},
  \bibinfo {author} {\bibfnamefont {B.~C.}\ \bibnamefont {Chakoumakos}},
  \bibinfo {author} {\bibfnamefont {J.~A.}\ \bibnamefont {Fernandez-Baca}},
  \bibinfo {author} {\bibfnamefont {H.~E.}\ \bibnamefont {Harmon}}, \bibinfo
  {author} {\bibfnamefont {J.~W.}\ \bibnamefont {Sharp}},\ and\ \bibinfo
  {author} {\bibfnamefont {E.~H.}\ \bibnamefont {Volckmann}},\ }\bibfield
  {title} {\bibinfo {title} {Magnetic, transport, and structural properties of
  {${\mathrm{Fe}}_{1\mathrm{\ensuremath{-}}\mathit{x}}$${\mathrm{Ir}}_{\mathit{x}}$Si}},\
  }\href {https://doi.org/10.1103/PhysRevB.50.8207} {\bibfield  {journal}
  {\bibinfo  {journal} {Phys. Rev. B}\ }\textbf {\bibinfo {volume} {50}},\
  \bibinfo {pages} {8207} (\bibinfo {year} {1994})}\BibitemShut {NoStop}%
\bibitem [{\citenamefont {Wartchow}\ \emph {et~al.}(1997)\citenamefont
  {Wartchow}, \citenamefont {Gerighausen},\ and\ \citenamefont
  {Binnewies}}]{Wartchow1997}%
  \BibitemOpen
  \bibfield  {author} {\bibinfo {author} {\bibfnamefont {R.}~\bibnamefont
  {Wartchow}}, \bibinfo {author} {\bibfnamefont {S.}~\bibnamefont
  {Gerighausen}},\ and\ \bibinfo {author} {\bibfnamefont {M.}~\bibnamefont
  {Binnewies}},\ }\bibfield  {title} {\bibinfo {title} {Redetermination of the
  crystal structure of iron silicide, {FeSi}},\ }\href
  {https://doi.org/doi:10.1524/ncrs.1997.212.1.320} {\bibfield  {journal}
  {\bibinfo  {journal} {Zeitschrift f\"ur Kristallographie - New Crystal
  Structures}\ }\textbf {\bibinfo {volume} {212}},\ \bibinfo {pages} {320}
  (\bibinfo {year} {1997})}\BibitemShut {NoStop}%
\bibitem [{\citenamefont {Zhao}\ \emph {et~al.}(2009)\citenamefont {Zhao},
  \citenamefont {Han}, \citenamefont {Yu}, \citenamefont {Xue},\ and\
  \citenamefont {Gao}}]{Zhao2009}%
  \BibitemOpen
  \bibfield  {author} {\bibinfo {author} {\bibfnamefont {Y.~N.}\ \bibnamefont
  {Zhao}}, \bibinfo {author} {\bibfnamefont {H.~L.}\ \bibnamefont {Han}},
  \bibinfo {author} {\bibfnamefont {Y.}~\bibnamefont {Yu}}, \bibinfo {author}
  {\bibfnamefont {W.~H.}\ \bibnamefont {Xue}},\ and\ \bibinfo {author}
  {\bibfnamefont {T.}~\bibnamefont {Gao}},\ }\bibfield  {title} {\bibinfo
  {title} {First-principles studies of the electronic and dynamical properties
  of monosilicides {MSi} ({M = Fe, Ru, Os})},\ }\href
  {https://doi.org/10.1209/0295-5075/85/47005} {\bibfield  {journal} {\bibinfo
  {journal} {{EPL} (Europhysics Letters)}\ }\textbf {\bibinfo {volume} {85}},\
  \bibinfo {pages} {47005} (\bibinfo {year} {2009})}\BibitemShut {NoStop}%
\bibitem [{\citenamefont {Arita}\ \emph {et~al.}(2008)\citenamefont {Arita},
  \citenamefont {Shimada}, \citenamefont {Takeda}, \citenamefont {Nakatake},
  \citenamefont {Namatame}, \citenamefont {Taniguchi}, \citenamefont {Negishi},
  \citenamefont {Oguchi}, \citenamefont {Saitoh}, \citenamefont {Fujimori},\
  and\ \citenamefont {Kanomata}}]{Arita2008}%
  \BibitemOpen
  \bibfield  {author} {\bibinfo {author} {\bibfnamefont {M.}~\bibnamefont
  {Arita}}, \bibinfo {author} {\bibfnamefont {K.}~\bibnamefont {Shimada}},
  \bibinfo {author} {\bibfnamefont {Y.}~\bibnamefont {Takeda}}, \bibinfo
  {author} {\bibfnamefont {M.}~\bibnamefont {Nakatake}}, \bibinfo {author}
  {\bibfnamefont {H.}~\bibnamefont {Namatame}}, \bibinfo {author}
  {\bibfnamefont {M.}~\bibnamefont {Taniguchi}}, \bibinfo {author}
  {\bibfnamefont {H.}~\bibnamefont {Negishi}}, \bibinfo {author} {\bibfnamefont
  {T.}~\bibnamefont {Oguchi}}, \bibinfo {author} {\bibfnamefont
  {T.}~\bibnamefont {Saitoh}}, \bibinfo {author} {\bibfnamefont
  {A.}~\bibnamefont {Fujimori}},\ and\ \bibinfo {author} {\bibfnamefont
  {T.}~\bibnamefont {Kanomata}},\ }\bibfield  {title} {\bibinfo {title}
  {Angle-resolved photoemission study of the strongly correlated semiconductor
  {FeSi}},\ }\href {https://doi.org/10.1103/PhysRevB.77.205117} {\bibfield
  {journal} {\bibinfo  {journal} {Phys. Rev. B}\ }\textbf {\bibinfo {volume}
  {77}},\ \bibinfo {pages} {205117} (\bibinfo {year} {2008})}\BibitemShut
  {NoStop}%
\bibitem [{\citenamefont {Klein}\ \emph {et~al.}(2009)\citenamefont {Klein},
  \citenamefont {Menzel}, \citenamefont {Doll}, \citenamefont {Neef},
  \citenamefont {Zur}, \citenamefont {Jursic}, \citenamefont {Schoenes},\ and\
  \citenamefont {Reinert}}]{Klein2009}%
  \BibitemOpen
  \bibfield  {author} {\bibinfo {author} {\bibfnamefont {M.}~\bibnamefont
  {Klein}}, \bibinfo {author} {\bibfnamefont {D.}~\bibnamefont {Menzel}},
  \bibinfo {author} {\bibfnamefont {K.}~\bibnamefont {Doll}}, \bibinfo {author}
  {\bibfnamefont {M.}~\bibnamefont {Neef}}, \bibinfo {author} {\bibfnamefont
  {D.}~\bibnamefont {Zur}}, \bibinfo {author} {\bibfnamefont {I.}~\bibnamefont
  {Jursic}}, \bibinfo {author} {\bibfnamefont {J.}~\bibnamefont {Schoenes}},\
  and\ \bibinfo {author} {\bibfnamefont {F.}~\bibnamefont {Reinert}},\
  }\bibfield  {title} {\bibinfo {title} {Photoemission spectroscopy across the
  semiconductor-to-metal transition in {FeSi}},\ }\href
  {https://doi.org/10.1088/1367-2630/11/2/023026} {\bibfield  {journal}
  {\bibinfo  {journal} {New Journal of Physics}\ }\textbf {\bibinfo {volume}
  {11}},\ \bibinfo {pages} {023026} (\bibinfo {year} {2009})}\BibitemShut
  {NoStop}%
\bibitem [{\citenamefont {Baroni}\ \emph {et~al.}(2010)\citenamefont {Baroni},
  \citenamefont {Giannozzi},\ and\ \citenamefont {Isaev}}]{Baroni2010}%
  \BibitemOpen
  \bibfield  {author} {\bibinfo {author} {\bibfnamefont {S.}~\bibnamefont
  {Baroni}}, \bibinfo {author} {\bibfnamefont {P.}~\bibnamefont {Giannozzi}},\
  and\ \bibinfo {author} {\bibfnamefont {E.}~\bibnamefont {Isaev}},\ }\bibfield
   {title} {\bibinfo {title} {{Density-Functional Perturbation Theory for
  Quasi-Harmonic Calculations}},\ }\href
  {https://doi.org/10.2138/rmg.2010.71.3} {\bibfield  {journal} {\bibinfo
  {journal} {Reviews in Mineralogy and Geochemistry}\ }\textbf {\bibinfo
  {volume} {71}},\ \bibinfo {pages} {39} (\bibinfo {year} {2010})}\BibitemShut
  {NoStop}%
\bibitem [{\citenamefont {Debernardi}\ \emph {et~al.}(2001)\citenamefont
  {Debernardi}, \citenamefont {Alouani},\ and\ \citenamefont
  {Dreyss\'e}}]{Debernardi2001}%
  \BibitemOpen
  \bibfield  {author} {\bibinfo {author} {\bibfnamefont {A.}~\bibnamefont
  {Debernardi}}, \bibinfo {author} {\bibfnamefont {M.}~\bibnamefont
  {Alouani}},\ and\ \bibinfo {author} {\bibfnamefont {H.}~\bibnamefont
  {Dreyss\'e}},\ }\bibfield  {title} {\bibinfo {title} {Ab initio
  thermodynamics of metals: Al and w},\ }\href
  {https://doi.org/10.1103/PhysRevB.63.064305} {\bibfield  {journal} {\bibinfo
  {journal} {Phys. Rev. B}\ }\textbf {\bibinfo {volume} {63}},\ \bibinfo
  {pages} {064305} (\bibinfo {year} {2001})}\BibitemShut {NoStop}%
\bibitem [{\citenamefont {Quong}\ and\ \citenamefont {Liu}(1997)}]{Quong1997}%
  \BibitemOpen
  \bibfield  {author} {\bibinfo {author} {\bibfnamefont {A.~A.}\ \bibnamefont
  {Quong}}\ and\ \bibinfo {author} {\bibfnamefont {A.~Y.}\ \bibnamefont
  {Liu}},\ }\bibfield  {title} {\bibinfo {title} {First-principles calculations
  of the thermal expansion of metals},\ }\href
  {https://doi.org/10.1103/PhysRevB.56.7767} {\bibfield  {journal} {\bibinfo
  {journal} {Phys. Rev. B}\ }\textbf {\bibinfo {volume} {56}},\ \bibinfo
  {pages} {7767} (\bibinfo {year} {1997})}\BibitemShut {NoStop}%
\bibitem [{\citenamefont {Reznik}\ and\ \citenamefont
  {Ahmadova}(2020)}]{Dmitry2020}%
  \BibitemOpen
  \bibfield  {author} {\bibinfo {author} {\bibfnamefont {D.}~\bibnamefont
  {Reznik}}\ and\ \bibinfo {author} {\bibfnamefont {I.}~\bibnamefont
  {Ahmadova}},\ }\bibfield  {title} {\bibinfo {title} {{Automating Analysis of
  Neutron Scattering Time-of-Flight Single Crystal Phonon Data}},\ }\href
  {https://doi.org/10.3390/qubs4040041} {\bibfield  {journal} {\bibinfo
  {journal} {Quantum Beam Science}\ }\textbf {\bibinfo {volume} {4}},\ \bibinfo
  {pages} {1} (\bibinfo {year} {2020})}\BibitemShut {NoStop}%
\bibitem [{\citenamefont {Racu}\ \emph {et~al.}(2007)\citenamefont {Racu},
  \citenamefont {Menzel}, \citenamefont {Schoenes},\ and\ \citenamefont
  {Doll}}]{Racu2007}%
  \BibitemOpen
  \bibfield  {author} {\bibinfo {author} {\bibfnamefont {A.-M.}\ \bibnamefont
  {Racu}}, \bibinfo {author} {\bibfnamefont {D.}~\bibnamefont {Menzel}},
  \bibinfo {author} {\bibfnamefont {J.}~\bibnamefont {Schoenes}},\ and\
  \bibinfo {author} {\bibfnamefont {K.}~\bibnamefont {Doll}},\ }\bibfield
  {title} {\bibinfo {title} {Crystallographic disorder and electron-phonon
  coupling in {${\mathrm{Fe}}_{1\ensuremath{-}x}{\mathrm{Co}}_{x}\mathrm{Si}$}
  single crystals: Raman spectroscopy study},\ }\href
  {https://doi.org/10.1103/PhysRevB.76.115103} {\bibfield  {journal} {\bibinfo
  {journal} {Phys. Rev. B}\ }\textbf {\bibinfo {volume} {76}},\ \bibinfo
  {pages} {115103} (\bibinfo {year} {2007})}\BibitemShut {NoStop}%
\bibitem [{\citenamefont {Rozenberg}\ \emph {et~al.}(1996)\citenamefont
  {Rozenberg}, \citenamefont {Kotliar},\ and\ \citenamefont
  {Kajueter}}]{Rozenberg1996}%
  \BibitemOpen
  \bibfield  {author} {\bibinfo {author} {\bibfnamefont {M.~J.}\ \bibnamefont
  {Rozenberg}}, \bibinfo {author} {\bibfnamefont {G.}~\bibnamefont {Kotliar}},\
  and\ \bibinfo {author} {\bibfnamefont {H.}~\bibnamefont {Kajueter}},\
  }\bibfield  {title} {\bibinfo {title} {Transfer of spectral weight in
  spectroscopies of correlated electron systems},\ }\href
  {https://doi.org/10.1103/PhysRevB.54.8452} {\bibfield  {journal} {\bibinfo
  {journal} {Phys. Rev. B}\ }\textbf {\bibinfo {volume} {54}},\ \bibinfo
  {pages} {8452} (\bibinfo {year} {1996})}\BibitemShut {NoStop}%
\end{thebibliography}%

\end{document}